\documentclass{article}%
\usepackage{amsfonts}
\usepackage{amssymb}
\usepackage{amsmath}
\usepackage{graphicx}%
\providecommand{\U}[1]{\protect\rule{.1in}{.1in}}

\newtheorem{proposition}{Proposition}

\begin{document}

\title{Counting Distinctions: \\On the Conceptual Foundations of Shannon's Information Theory}
\author{David Ellerman\thanks{This paper is dedicated to the memory of Gian-Carlo
Rota---mathematician, philosopher, mentor, and friend.}\\Department of Philosophy\\University of California at Riverside}
\maketitle
\tableofcontents

\begin{abstract}
Categorical logic has shown that modern logic is essentially the logic of
subsets (or "subobjects"). Partitions are dual to subsets so there is a dual
logic of partitions where a "distinction" [an ordered pair of distinct
elements $(u,u^{\prime})$ from the universe $U$ ] is dual to an "element". An
element being in a subset is analogous to a partition $\pi$ on $U$ making a
distinction, i.e., if $u$ and $u^{\prime}$ were in different blocks of $\pi$.
Subset logic leads to finite probability theory by taking the (Laplacian)
probability as the normalized size of each subset-event of a finite universe.
The analogous step in the logic of partitions is to assign to a partition the
number of distinctions made by a partition normalized by the total number of
ordered pairs $|U|%
{{}^2}%
$ from the finite universe. That yields a notion of "logical entropy" for
partitions and a "logical information theory." The logical theory directly
counts the (normalized) number of distinctions in a partition while Shannon's
theory gives the average number of binary partitions needed to make those same
distinctions. Thus the logical theory is seen as providing a conceptual
underpinning for Shannon's theory based on the logical notion of "distinctions."

\end{abstract}

\section{Towards a Logic of Partitions}

In ordinary logic, a statement $P\left(  a\right)  $ is formed by a predicate
$P\left(  x\right)  $ applying to an individual name \textquotedblleft%
$a$\textquotedblright\ (which could be an $n$-tuple in the case of relations).
The predicate is modeled by a subset $S_{P}$ of a universe set $U$ and an
individual name such as \textquotedblleft$a$\textquotedblright\ would be
assigned an individual $u_{a}\in U$ (an $n$-tuple in the case of relations).
The statement $P\left(  a\right)  $ would hold in the model if $u_{a}\in
S_{P}$. In short, logic is modeled as the \textit{logic of subsets} of a set.
Largely due to the efforts of William Lawvere, the modern treatment of logic
was reformulated and vastly generalized using category theory in what is now
called \textit{categorical logic}. Subsets were generalized to subobjects or
\textquotedblleft parts\textquotedblright\ (equivalence classes of
monomorphisms) so that logic has become the logic of subobjects.\footnote{See
\cite{law:sfm} Appendix A for a good treatment.}

There is a duality between \textit{subsets} of a set and \textit{partitions}%
\footnote{A \textit{partition} $\pi$ on a set $U$ is usually defined as a
mutually exclusive and jointly exhaustive set $\left\{  B\right\}  _{B\in\pi}$
of subsets or \textquotedblleft blocks\textquotedblright\ $B\subseteq U$.
Every equivalence relation on a set $U$ determines a partition on $U$ (with
the equivalence classes as the blocks) and vice-versa. For our purposes, it is
useful to think of partitions as binary relations defined as the complement to
an equivalence relation in the set of ordered pairs $U\times U$. Intuitively,
they have complementary functions in the sense that equivalence relations
identify while partitions distinguish elements of $U$.} on a set.
\textquotedblleft The dual notion (obtained by reversing the arrows) of `part'
is the notion of \textit{partition}.\textquotedblright\cite[p. 85]{law:sfm} In
category theory, this emerges as the reverse-the-arrows duality between
monomorphisms (monos), e.g., injective set functions, and epimorphisms (epis),
e.g., surjective set functions, and between subobjects and quotient objects.
If modern logic is formulated as the logic of subsets, or more generally,
subobjects or \textquotedblleft parts\textquotedblright, then the question
naturally arises of a dual logic that might play the analogous role for
partitions and their generalizations.

Quite aside from category theory duality, it has long been noted in
combinatorial mathematics, e.g., in Gian-Carlo Rota's work in combinatorial
theory and probability theory \cite{Rota:Prob79}, that there is a type of
duality between subsets of a set and partitions on a set. Just as subsets of a
set are partially ordered by inclusion, so partitions on a set are partially
ordered by refinement.\footnote{A partition $\pi$ more refined than a
partition $\sigma$, written $\sigma\preceq\pi$, if each block of $\pi$ is
contained in some block of $\sigma$. Much of the older literature (e.g.,
\cite[Example 6, p. 2]{birk:lt}) writes this relationship the other way around
but, for reasons that will become clear, we are adopting a newer way of
writing refinement (e.g., \cite{Gray:eit}) so that the more refined partition
is higher in the refinement ordering.} Moreover, both partial orderings are in
fact lattices (i.e., have meets and joins) with a top element $\widehat{1}$
and a bottom element $\widehat{0}$. In the lattice of all subsets
$\mathcal{P}(U)$ (the power set) of a set $U$, the meet and join are, of
course, intersection and union while the top element is the universe $U$ and
the bottom element is the null set $\emptyset$. In the lattice of all
partitions $\Pi(U)$ on a non-empty set $U$, there are also meet and join
operations (defined later) while the bottom element is the indiscrete
partition (the \textquotedblleft blob\textquotedblright) where all of $U$ is
one block and the top element is the discrete partition where each element of
$U$ is a singleton block.\footnote{Rota and his students have developed a
logic for a special type of equivalence relation (which is rather ubiquitous
in mathematics) using join and meet as the only connectives.\cite{fmr:cer}}

This paper is part of a research programme to develop the general dual logic
of partitions. The principal novelty in this paper is an analogy between the
usual semantics for subset logic and a suggested semantics for partition
logic; the themes of the paper unfold from that starting point. Starting with
the analogy between a subset of a set and a partition on the set, the analogue
to the notion of an \textit{element} of a subset is the notion of a
\textit{distinction} of a partition which is simply an ordered pair $\left(
u,u^{\prime}\right)  \in U\times U$ in distinct blocks of the
partition.\footnote{Intuitively we might think of an element of a set as an
\textquotedblleft it.\textquotedblright\ We will argue that a distinction or
\textquotedblleft dit\textquotedblright\ is the corresponding logical atom of
information.\ In economics, there is a basic distinction between rivalrous
goods (where more for one means less for another) such a material things
(\textquotedblleft its\textquotedblright) in contrast to non-rivalrous goods
(where what one person acquires does not take away from another) such as
ideas, knowledge, and information (\textquotedblleft bits\textquotedblright%
\ or \textquotedblleft dits\textquotedblright). In that spirit, an element of
a set represents a material thing, an \textquotedblleft it,\textquotedblright%
\ while the dual notion of a distinction or \textquotedblleft
dit\textquotedblright\ represents the immaterial notion of two
\textquotedblleft its\textquotedblright\ being distinct. The distinction
between $u$ and $u^{\prime}$ is the fact that $u\not =u^{\prime}$, not a new
\textquotedblleft thing\textquotedblright\ or \textquotedblleft
it.\textquotedblright\ But for mathematical purposes we may represent a
distinction by a pair of distinct elements such as the ordered pair $\left(
u,u^{\prime}\right)  $ which is a higher level \textquotedblleft
it,\textquotedblright\ i.e., an element in the Cartesian product of a set with
itself (see next section).} The logic of subsets leads to finite probability
theory where events are subsets $S$ of a finite sample space $U$ and which
assigns probabilities $\operatorname*{Prob}\left(  S\right)  $ to subsets
(e.g., the Laplacian equiprobable distribution where $\operatorname*{Prob}%
\left(  S\right)  =\left\vert S\right\vert /\left\vert U\right\vert $).
Following the suggested analogies, the logic of partitions similarly leads to
a \textquotedblleft logical\textquotedblright\ information theory where the
numerical value naturally assigned to a partition can be seen as the
\textit{logical information content} or \textit{logical entropy} $h\left(
\pi\right)  $ of the partition. It is initially defined in a Laplacian manner
as the number of distinctions that a partition makes normalized by the number
of ordered pairs of the universe set $U$. The probability interpretation of
$h\left(  \pi\right)  $ is the probability that a random pair from $U\times U$
is distinguished by $\pi$, just as $\operatorname*{Prob}\left(  S\right)  $ is
the probability that a random choice from $U$ is an element of $S$. This
logical entropy is precisely related to Shannon's entropy measure\textit{\ }%
\cite{Shannon:mtc} so the development of logical information theory can be
seen as providing a new conceptual basis for information theory at the basic
level of logic using \textquotedblleft distinctions\textquotedblright\ as the
conceptual atoms.

Historically and conceptually, probability theory started with the simple
logical operations on subsets (e.g., union, intersection, and complementation)
and assigned a numerical measure to subsets of a finite set of outcomes
(number of favorable outcomes divided by\ the total number of outcomes). Then
probability theory \textquotedblleft took off\textquotedblright\ from these
simple beginnings to become a major branch of pure and applied mathematics.

The research programme for partition logic that underlies this paper sees
Shannon's information theory as \textquotedblleft taking off\textquotedblright%
\ from the simple notions of partition logic in analogy with the conceptual
development of probability theory that starts with simple notions of subset
logic. But historically, Shannon's information theory appeared "as a bolt out
of the blue" in a rather sophisticated and axiomatic form. Moreover, partition
logic is still in its infancy today, not to mention the over half a century
ago when Shannon's theory was published.\footnote{For instance, the conceptual
beginnings of probability theory in subset logic is shown by the role of
Boolean algebras in probability theory, but what is the corresponding algebra
for partition logic?} But starting with the suggested semantics for partition
logic (i.e., the subset-to-partition and element-to-distinction analogies), we
develop the partition analogue (\textquotedblleft counting
distinctions\textquotedblright) of the beginnings of finite probability theory
(\textquotedblleft counting outcomes\textquotedblright), and then we show how
it is related to the already-developed information theory of Shannon. It is in
that sense that the developments in the paper provide a logical or conceptual
foundation (\textquotedblleft foundation\textquotedblright\ in the sense of a
basic conceptual starting point) for information theory.\footnote{Perhaps an
analogy will be helpful. It is \textit{as if} the axioms for probability
theory had first emerged full-blown from Kolmogorov \cite{kol:prob} and then
one realized belatedly that the discipline could be seen as growing out of the
starting point of operations on subsets of a finite space of outcomes where
the logic was the logic of subsets.}

The following table sets out some of the analogies in a concise form (where
the diagonal in $U\times U$ is $\Delta_{U}=\left\{  \left(  u,u\right)  |u\in
U\right\}  $).

\begin{center}%
\begin{tabular}
[c]{l|l|l|}\cline{2-3}%
\textit{Table of Analogies} & \textbf{Subsets} & \textbf{Partitions}\\\hline
\multicolumn{1}{|l|}{\textquotedblleft Atoms\textquotedblright} & Elements &
Distinctions\\\hline
\multicolumn{1}{|l|}{All atoms} & Universe $U$ (all $u\in U$) = $\widehat{1}$
& Discrete partition $\widehat{1}$ (all dits)\\\hline
\multicolumn{1}{|l|}{No atoms} & Null set $\emptyset$ (no $u\in U$) =
$\widehat{0}$ & Indiscrete partition $\widehat{0}$ (no dits)\\\hline
\multicolumn{1}{|l|}{Model of proposition or event} & Subset $S\subseteq U$ &
Partition $\pi$ on $U$\\\hline
\multicolumn{1}{|l|}{Model of individual or outcome} & Element $u$ in $U$ &
Distinction $\left(  u,u^{\prime}\right)  $ in $U\times U-\Delta_{U}$\\\hline
\multicolumn{1}{|l|}{Prop. holds or event occurs} & Element $u$ in subset $S$
& Partition $\pi$ distinguishes $\left(  u,u^{\prime}\right)  $\\\hline
\multicolumn{1}{|l|}{Lattice of propositions/events} & Lattice of all subsets
$\mathcal{P}\left(  U\right)  $ & Lattice of all partitions $\Pi\left(
U\right)  $\\\hline
\multicolumn{1}{|l|}{Counting measure ($U$ finite)} & \# elements in $S$ & \#
dits (as ordered pairs) in $\pi$\\\hline
\multicolumn{1}{|l|}{Normalized count ($U$ finite)} & $\operatorname*{Prob}%
(S)=\frac{\text{\# elements in }S}{\left\vert U\right\vert }$ & $h\left(
\pi\right)  =$ $\frac{\text{\#distinctions in }\pi}{\left\vert U\times
U\right\vert }$\\\hline
\multicolumn{1}{|l|}{Prob. Interpretation ($U$ finite)} &
$\operatorname*{Prob}\left(  S\right)  =$ probability that & $h\left(
\pi\right)  =$ probability random pair\\
\multicolumn{1}{|l|}{} & random element $u$ is in $S$ & $\left(  u,u^{\prime
}\right)  $ is distinguished by $\pi$\\\hline
\end{tabular}

\end{center}

These analogies show one set of reasons why the lattice of partitions
$\Pi\left(  U\right)  $ should be written with the discrete partition as the
top element and the indiscrete partition (blob) as the bottom element of the
lattice---in spite of the usual convention of writing the \textquotedblleft
refinement\textquotedblright\ ordering the other way around as what Gian-Carlo
Rota called the \textquotedblleft unrefinement ordering.\textquotedblright

With this motivation, we turn to the development of this conceptual basis for
information theory.

\section{Logical Information Theory}

\subsection{The Closure Space $U\times U$}

Claude Shannon's classic 1948 articles \cite{Shannon:mtc} developed a
statistical theory of communications that is ordinarily called
\textquotedblleft information theory.\textquotedblright\ Shannon built upon
the work of Ralph Hartley \cite{hart:ti} twenty years earlier. After Shannon's
information theory was presented axiomatically, there was a spate of new
definitions of \textquotedblleft entropy\textquotedblright\ with various
axiomatic properties but without concrete (never mind logical) interpretations
\cite{kapur:mi}. Here we take the approach of starting with a notion that
arises naturally in the logic of partitions, dual to the usual logic of
subsets. The notion of a distinction or \textquotedblleft
dit\textquotedblright\ is taken as the logical atom of information and a
\textquotedblleft logical information theory\textquotedblright\ is developed
based on that interpretation. When the universe set $U$ is finite, then we
have a numerical notion of \textquotedblleft information\textquotedblright\ or
\textquotedblleft entropy\textquotedblright\ $h\left(  \pi\right)  $ of a
partition $\pi$ in the number of distinctions normalized by the number of
ordered pairs. This logical \textquotedblleft counting
distinctions\textquotedblright\ notion of information or entropy can then be
related to Shannon's measure of information or entropy.

The basic conceptual unit in logical information theory is the distinction or
\textit{dit} (from \textquotedblleft DIsTinction\textquotedblright\ but
motivated by \textquotedblleft bit\textquotedblright). A pair $\left(
u,u^{\prime}\right)  $ of distinct elements of $U$ are distinguished by $\pi$,
i.e., form a dit of $\pi$, if $u$ and $u^{\prime}$ are in different blocks of
$\pi$.\footnote{One might also develop the theory using unordered pairs
$\left\{  u,u^{\prime}\right\}  $ but the later development of the theory
using probabilistic methods is much facilitated by using ordered pairs
$\left(  u,u^{\prime}\right)  $. Thus for $u\not =u^{\prime}$, $\left(
u,u^{\prime}\right)  $ and $\left(  u^{\prime},u\right)  $ count as two
distinctions. This means that the count of distinctions in a partition must be
normalized by $\left\vert U\times U\right\vert $. Note that $U\times U$
includes the diagonal self-pairs $\left(  u,u\right)  $ which can never be
distinctions.} A pair $\left(  u,u^{\prime}\right)  $ are identified by $\pi$
and form an \textit{indit }(from INDIsTinction or \textquotedblleft
identification\textquotedblright) of the partition if they are contained in
the same block of $\pi$. A partition on $U$ can be characterized by either its
dits or indits (just as a subset $S$ of $U$ can be characterized by the
elements added to the null set to arrive at $S$ or by the elements of $U$
thrown out to arrive at $S$). When a partition $\pi$ is thought of as
determining an equivalence relation, then the equivalence relation, as a set
of ordered pairs contained in $U\times U=U^{2}$, is the \textit{indit set}
$\operatorname*{indit}(\pi)$ of indits of the partition. But from the view
point of logical information theory, the focus is on the distinctions, so the
partition $\pi$ qua binary relation is given by the complementary \textit{dit
set} $\operatorname*{dit}\left(  \pi\right)  $ of dits where
$\operatorname*{dit}\left(  \pi\right)  =\left(  U\times U\right)
-\operatorname*{indit}\left(  \pi\right)  =\operatorname*{indit}\left(
\pi\right)  ^{c}$. Rather than think of the partition as resulting from
identifications made to the elements of $U$ (i.e., distinctions excluded from
the discrete partition), we think of it as being formed by making distinctions
starting with the blob. This is analogous to a subset $S$ being thought of as
the set of elements that must be added to the null set to obtain $S$ rather
than the complementary approach to $S$ by giving the elements excluded from
$U$ to arrive at $S$. From this viewpoint, the natural ordering $\sigma
\preceq\pi$ of partitions would be given by the inclusion ordering of dit-sets
$\operatorname*{dit}\left(  \sigma\right)  \subseteq\operatorname*{dit}\left(
\pi\right)  $ and that is exactly the new way of writing the refinement
relation that we are using, i.e.,

\begin{center}
$\sigma\preceq\pi$ iff $\operatorname*{dit}\left(  \sigma\right)
\subseteq\operatorname*{dit}\left(  \pi\right)  $.
\end{center}

There is a natural (\textquotedblleft built-in\textquotedblright) closure
operation on $U\times U$ so that the equivalence relations on $U$ are given
(as binary relations) by the closed sets. A subset $C\subseteq U^{2}$ is
\textit{closed} if it contains the diagonal $\left\{  \left(  u,u\right)  \mid
u\in U\right\}  $, if $\left(  u,u^{\prime}\right)  \in C$ implies $\left(
u^{\prime},u\right)  \in C$, and if $\left(  u,u^{\prime}\right)  $ and
$\left(  u^{\prime},u^{\prime\prime}\right)  $ are in $C$, then $\left(
u,u^{\prime\prime}\right)  $ is in $C$. Thus the closed sets of $U^{2}$ are
the reflexive, symmetric, and transitive relations, i.e., the equivalence
relations on $U$. The intersection of closed sets is closed and the
intersection of all closed sets containing a subset $S\subseteq U^{2}$ is the
\textit{closure} $\overline{S}$ of $S$.

It should be carefully noted that the closure operation on the closure space
$U^{2}$ is not a \textit{topological} closure operation in the sense that the
union of two closed set is not necessarily closed. In spite of the closure
operation not being topological, we may still refer to the complements of
closed sets as being \textit{open} sets, i.e., the dit sets of partitions on
$U$. As usual, the \textit{interior} $\operatorname*{int}(S)$ of any subset
$S$ is defined as the complement of the closure of its complement:
$\operatorname*{int}(S)=\left(  \overline{S^{c}}\right)  ^{c}$.

The open sets of $U\times U$ ordered by inclusion form a lattice isomorphic to
the lattice $\Pi(U)$ of partitions on $U$. The closed sets of $U\times U$
ordered by inclusion form a lattice isomorphic to $\Pi(U)^{op}$, the opposite
of the lattice of partitions on $U$ (formed by turning around the partial
order). The motivation for writing the refinement relation in the old way was
probably that equivalence relations were thought of as binary relations
$\operatorname*{indit}\left(  \pi\right)  \subseteq U\times U$, so the
ordering of equivalence relations was written to reflect the inclusion
ordering between indit-sets. But since a partition and an equivalence relation
were then taken as essentially the \textquotedblleft same
thing,\textquotedblright\ i.e., a set $\left\{  B\right\}  _{B\in\pi}$ of
mutually exclusive and jointly exhaustive subsets (\textquotedblleft
blocks\textquotedblright\ or \textquotedblleft equivalence
classes\textquotedblright) of $U$, that way of writing the ordering carried
over to partitions. But we identify a partition $\pi$ \textit{as a binary
relation} with its dit-set $\operatorname*{dit}\left(  \pi\right)  =U\times
U-\operatorname*{indit}\left(  \pi\right)  $ so our refinement ordering is the
inclusion ordering between dit-sets (the opposite of the inclusion ordering of
indit-sets).\footnote{One way to establish the duality between elements of
subsets and distinctions in a partition is to start with the refinement
relation as the partial order in the lattice of partitions $\Pi(U)$ analogous
to the inclusion partial order in the lattice of subsets $\mathcal{P}(U)$.
Then the mapping $\pi\longmapsto\operatorname*{dit}(\pi)$ represents the
lattice of partitions as the lattice of open subsets of the closure space
$U\times U$ with inclusion as the partial order. Then the analogue of the
elements in the subsets of $\mathcal{P}(U)$ would be the elements in the
subsets $\operatorname*{dit}\left(  \pi\right)  $ representing the partitions,
namely, the distinctions.}

Given two partitions $\pi$ and $\sigma$ on $U$, the open set corresponding to
the \textit{join} $\pi\vee\sigma$ of the partitions is the partition whose
dit-set is the union of their dit-sets:\footnote{Note that this union of dit
sets gives the dit set of the \textquotedblleft meet\textquotedblright\ in the
old reversed way of writing the refinement ordering.}

\begin{center}
$\operatorname*{dit}(\pi\vee\sigma)=\operatorname*{dit}\left(  \pi\right)
\cup\operatorname*{dit}\left(  \sigma\right)  $.
\end{center}

\noindent The open set corresponding to the \textit{meet} $\pi\wedge\sigma$ of
partitions is the interior of the intersection of their
dit-sets:\footnote{Note that this is the \textquotedblleft
join\textquotedblright\ in the old reversed way of writing the refinement
ordering. This operation defined by the interior operator of the
non-topological closure operation leads to \textquotedblleft
anomolous\textquotedblright\ results such as the non-distributivity of the
partition lattice---in contrast to the distributivity of the lattice of open
sets of a topological space.}

\begin{center}
$\operatorname*{dit}(\pi\wedge\sigma)=\operatorname*{int}\left(
\operatorname*{dit}\left(  \pi\right)  \cap\operatorname*{dit}\left(
\sigma\right)  \right)  $.
\end{center}

\noindent The open set corresponding to the bottom or blob $\widehat{0}$ is
the null set $\emptyset\subseteq$ $U\times U$ (no distinctions) and the open
set corresponding to the discrete partition or top $\widehat{1}$ is the
complement of the diagonal, $U\times U-\Delta_{U}$ (all distinctions).

\subsection{Some Set Structure Theorems}

Before restricting ourselves to finite $U$ to use the counting measure
$\left\vert \operatorname*{dit}\left(  \pi\right)  \right\vert $, there are a
few structure theorems that are independent of cardinality. If the
\textquotedblleft atom\textquotedblright\ of information is the dit then the
atomic information in a partition $\pi$ \textquotedblleft is\textquotedblright%
\ its dit set, $\operatorname*{dit}(\pi)$. The information common to two
partitions $\pi$ and $\sigma$, their \textit{mutual information set}, would
naturally be the intersection of their dit sets (which is not necessarily the
dit set of a partition):

\begin{center}
$\operatorname*{Mut}(\pi,\sigma)=\operatorname*{dit}\left(  \pi\right)
\cap\operatorname*{dit}\left(  \sigma\right)  $.
\end{center}

\noindent Shannon deliberately defined his measure of information so that it
would be \textquotedblleft additive\textquotedblright\ in the sense that the
measure of information in two independent probability distributions would be
the sum of the information measures of the two separate distributions and
there would be zero mutual information between the independent distributions.
But this is not true at the logical level with information defined as
distinctions. There is \textit{always} mutual information between two non-blob
partitions---even though the interior of $\operatorname*{Mut}\left(
\pi,\sigma\right)  $ might be empty, i.e., $\operatorname*{int}\left(
\operatorname*{Mut}(\pi,\sigma\right)  )=\operatorname*{int}\left(
\operatorname*{dit}\left(  \pi\right)  \cap\operatorname*{dit}\left(
\sigma\right)  \right)  =\operatorname*{dit}\left(  \pi\wedge\sigma\right)  $
might be empty so that $\pi\wedge\sigma=\widehat{0}$.

\begin{proposition}
Given two partitions $\pi$ and $\sigma$ on $U$ with $\pi\neq\widehat{0}%
\neq\sigma$, $\operatorname*{Mut}\left(  \pi,\sigma\right)  \neq\emptyset
$.\footnote{The contrapositive of this proposition is interesting. Given two
equivalence relations $E_{1},E_{2}\subseteq U^{2}$, if every pair of elements
$u,u^{\prime}\in U$ is identified by one or the other of the relations, i.e.,
$E_{1}\cup E_{2}=U^{2}$, then either $E_{1}=U^{2}$ or $E_{2}=U^{2}$.}
\end{proposition}

\noindent Since $\pi$ is not the blob, consider two elements $u$ and
$u^{\prime}$ distinguished by $\pi$ but identified by $\sigma$ [otherwise
$\left(  u,u^{\prime}\right)  \in\operatorname*{Mut}(\pi,\sigma)$]. Since
$\sigma$ is also not the blob, there must be a third element $u^{\prime\prime
}$ not in the same block of $\sigma$ as $u$ and $u^{\prime}$. But since $u$
and $u^{\prime}$ are in different blocks of $\pi$, the third element
$u^{\prime\prime}$ must be distinguished from one or the other or both in
$\pi$. Hence $\left(  u,u^{\prime\prime}\right)  $ or $\left(  u^{\prime
},u^{\prime\prime}\right)  $ must be distinguished by both partitions and thus
must be in their mutual information set $\operatorname*{Mut}\left(  \pi
,\sigma\right)  $.$\blacksquare$ (= end of proof marker)

The closed and open subsets of $U^{2}$ can be characterized using the usual
notions of blocks of a partition. Given a partition $\pi$ on $U$ as a set of
blocks $\pi=\left\{  B\right\}  _{B\in\pi}$, let $B\times B^{\prime}$ be the
Cartesian product of $B$ and $B^{\prime}$. Then%

\begin{align*}
\operatorname*{indit}\left(  \pi\right)   &  =%
{\textstyle\bigcup\limits_{B\in\pi}}
B\times B\\
\operatorname*{dit}\left(  \pi\right)   &  =%
{\textstyle\bigcup\limits_{\substack{B\neq B^{\prime} \\B,B^{\prime}\in\pi}}}
B\times B^{\prime}=U\times U-\operatorname*{indit}\left(  \pi\right)
=\operatorname*{indit}\left(  \pi\right)  ^{c}.
\end{align*}

\noindent The mutual information set can also be characterized in this manner.

\begin{proposition}
Given partitions $\pi$ and $\sigma$ with blocks $\left\{  B\right\}  _{B\in
\pi}$ and $\left\{  C\right\}  _{C\in\sigma}$, then
\end{proposition}

\begin{center}
$\operatorname*{Mut}\left(  \pi,\sigma\right)  =%
{\textstyle\bigcup\limits_{B\in\pi,C\in\sigma}}
\left(  B-\left(  B\cap C\right)  \right)  \times\left(  C-\left(  B\cap
C\right)  \right)  =\bigcup\limits_{B\in\pi,C\in\sigma}\left(  B-C\right)
\times\left(  C-B\right)  $.
\end{center}

\noindent The union (which is a disjoint union) will include the pairs
$\left(  u,u^{\prime}\right)  $ where for some $B\in\pi$ and $C\in\sigma$,
$u\in B-\left(  B\cap C\right)  $ and $u^{\prime}\in C-\left(  B\cap C\right)
$. Since $u^{\prime}$ is in $C$ but not in the intersection $B\cap C$, it must
be in a different block of $\pi$ than $B$ so $\left(  u,u^{\prime}\right)
\in\operatorname*{dit}\left(  \pi\right)  $. Symmetrically, $\left(
u,u^{\prime}\right)  \in\operatorname*{dit}\left(  \sigma\right)  $ so
$\left(  u,u^{\prime}\right)  \in\operatorname*{Mut}\left(  \pi,\sigma\right)
=\operatorname*{dit}\left(  \pi\right)  \cap\operatorname*{dit}\left(
\sigma\right)  $. Conversely if $\left(  u,u^{\prime}\right)  \in
\operatorname*{Mut}\left(  \pi,\sigma\right)  $ then take the $B$ containing
$u$ and the $C$ containing $u^{\prime}$. Since $\left(  u,u^{\prime}\right)  $
is distinguished by both partitions, $u\not \in C$ and $u^{\prime}\not \in B$
so that $\left(  u,u^{\prime}\right)  \in\left(  B-\left(  B\cap C\right)
\right)  \times\left(  C-\left(  B\cap C\right)  \right)  $.$\blacksquare$

\subsection{Logical Information Theory on Finite Sets}

For a finite set $U$, the (normalized) \textquotedblleft counting
distinctions\textquotedblright\ measure of information can be defined and
compared to Shannon's measure for finite probability distributions. Since the
information set of a partition $\pi$ on $U$ is its set of distinctions
$\operatorname*{dit}\left(  \pi\right)  $, the un-normalized numerical measure
of the information of a partition is simply the count of that set, $\left\vert
\operatorname*{dit}\left(  \pi\right)  \right\vert $ (\textquotedblleft dit
count\textquotedblright). But to account for the total number of ordered pairs
of elements from $U$, we normalize by $\left\vert U\times U\right\vert
=\left\vert U\right\vert ^{2}$ to obtain the \textit{logical information
content} or \textit{logical entropy} of a partition $\pi$ as its normalized
dit count:

\begin{center}
\fbox{$h\left(  \pi\right)  =\frac{\left\vert \operatorname*{dit}\left(
\pi\right)  \right\vert }{\left\vert U\times U\right\vert }$}.
\end{center}

Probability theory started with the finite case where there was a finite set
$U$ of possibilities (the finite sample space) and an event was a subset
$S\subseteq U$. Under the Laplacian assumption that each outcome was
equiprobable, the probability of the event $S$ was the similar normalized
counting measure of the set:

\begin{center}
$\operatorname*{Prob}\left(  S\right)  =\frac{\left\vert S\right\vert
}{\left\vert U\right\vert }$.
\end{center}

\noindent This is the probability that any randomly chosen element of $U$ is
an element of the subset $S$. In view of the dual relationship between being
in a subset and being distinguished by a partition, the analogous concept
would be the probability that an ordered pair $\left(  u,u^{\prime}\right)  $
of elements of $U$ chosen independently (i.e., with
replacement\footnote{Drawing with replacement would allow diagonal pairs
$\left(  u,u\right)  $ to be drawn and requires $\left\vert U\times
U\right\vert $ as the normalizing factor.}) would be distinguished by a
partition $\pi$, and that is precisely the logical entropy $h(\pi)=\left\vert
\operatorname*{dit}\left(  \pi\right)  \right\vert /\left\vert U\times
U\right\vert $ (since each pair randomly chosen from $U\times U$ is equiprobable).

\begin{center}
\fbox{Probabilistic interpretation: $h\left(  \pi\right)  =$ probability a
random pair is distinguished by $\pi$}.
\end{center}

\noindent In finite probability theory, when a point is sampled from the
sample space $U$, we say the event $S$ \textit{occurs} if the point $u$ was an
element in $S\subseteq U$. When a random pair $\left(  u,u^{\prime}\right)  $
is sampled from the sample space $U\times U$, we say the partition $\pi$
\textit{distinguishes\footnote{Equivalent terminology would be
\textquotedblleft differentiates\textquotedblright\ or \textquotedblleft
discriminates.\textquotedblright}} if the pair is distinguished by the
partition, i.e., if $\left(  u,u^{\prime}\right)  \in\operatorname*{dit}%
\left(  \pi\right)  \subseteq U\times U$. Then just as we take
$\operatorname*{Prob}\left(  S\right)  $ as the probability that the event $S$
occurs, so the logical entropy $h\left(  \pi\right)  $ is the probability that
the partition $\pi$ distinguishes.

Since $\operatorname*{dit}\left(  \pi\vee\sigma\right)  =\operatorname*{dit}%
\left(  \pi\right)  \cup\operatorname*{dit}\left(  \sigma\right)  $,

\begin{center}
\fbox{probability that $\pi\vee\sigma$ distinguishes $=h\left(  \pi\vee
\sigma\right)  =$ probability that $\pi$ or $\sigma$ distinguishes}.
\end{center}

The probability that a randomly chosen pair would be distinguished by $\pi$
\textit{and} $\sigma$ would be given by the relative cardinality of the mutual
information set which is called the \textit{mutual information} of the partitions:

\begin{center}
\fbox{Mutual logical information: $m(\pi,\sigma)=$ $\frac{\left\vert
\operatorname*{Mut}\left(  \pi,\sigma\right)  \right\vert }{\left\vert
U\right\vert ^{2}}=$ probability that $\pi$ and $\sigma$ distinguishes }.
\end{center}

Since the cardinality of intersections of sets can be analyzed using the
inclusion-exclusion principle, we have:

\begin{center}
$\left\vert \operatorname*{Mut}\left(  \pi,\sigma\right)  \right\vert
=\left\vert \operatorname*{dit}\left(  \pi\right)  \cap\operatorname*{dit}%
\left(  \sigma\right)  \right\vert =\left\vert \operatorname*{dit}\left(
\pi\right)  \right\vert +\left\vert \operatorname*{dit}\left(  \sigma\right)
\right\vert -\left\vert \operatorname*{dit}\left(  \pi\right)  \cup
\operatorname*{dit}\left(  \sigma\right)  \right\vert $.
\end{center}

\noindent Normalizing, the probability that a random pair is distinguished by
both partitions is given by the modular law:

\begin{center}
$m\left(  \pi,\sigma\right)  =\frac{\left\vert \operatorname*{dit}\left(
\pi\right)  \cap\operatorname*{dit}\left(  \sigma\right)  \right\vert
}{\left\vert U\right\vert ^{2}}=\frac{\left\vert \operatorname*{dit}\left(
\pi\right)  \right\vert }{\left\vert U\right\vert ^{2}}+\frac{\left\vert
\operatorname*{dit}\left(  \sigma\right)  \right\vert }{\left\vert
U\right\vert ^{2}}-\frac{\left\vert \operatorname*{dit}\left(  \pi\right)
\cup\operatorname*{dit}\left(  \sigma\right)  \right\vert }{\left\vert
U\right\vert ^{2}}=h\left(  \pi\right)  +h\left(  \sigma\right)  -h\left(
\pi\vee\sigma\right)  $.
\end{center}

\noindent This can be extended by the inclusion-exclusion principle to any
number of partitions. The mutual information set $\operatorname*{Mut}\left(
\pi,\sigma\right)  $ is not the dit-set of a partition but its interior is the
dit-set of the meet so the logical entropies of the join and meet satisfy the:

\begin{center}
Submodular inequality: $h\left(  \pi\wedge\sigma\right)  +h\left(  \pi
\vee\sigma\right)  \leq h\left(  \pi\right)  +h\left(  \sigma\right)  $.
\end{center}

\subsection{Using General Finite Probability Distributions}

Since the logical entropy of a partition on a finite set can be given a simple
probabilistic interpretation, it is not surprising that many methods of
probability theory can be harnessed to develop the theory. The theory for the
finite case can be developed at two different levels of generality, using the
specific Laplacian equiprobability distribution on the finite set $U$ or using
an arbitrary finite probability distribution. Correctly formulated, all the
formulas concerning logical entropy and the related concepts will work for the
general case, but our purpose is not mathematical generality. Our purpose is
to give the basic motivating example of logical entropy based on
\textquotedblleft counting distinctions\textquotedblright\ and to show its
relationship to Shannon's notion of entropy, thereby clarifying the logical
foundations of the latter concept.

Every probability distribution on a finite set $U$ gives a probability $p_{B}
$ for each block $B$ in a partition $\pi$ but for the Laplacian distribution,
it is just the relative cardinality of the block: $p_{B}=\frac{\left\vert
B\right\vert }{\left\vert U\right\vert }$ for blocks $B\in\pi$. Since there
are no empty blocks, $p_{B}>0$ and $\sum_{B\in\pi}p_{B}=1$. Since the dit set
of a partition is $\operatorname*{dit}\left(  \pi\right)  =%
{\textstyle\bigcup\limits_{B\neq B^{\prime}}}
B\times B^{\prime}$, its size is $\left\vert \operatorname*{dit}\left(
\pi\right)  \right\vert =\sum_{B\not =B^{\prime}}\left\vert B\right\vert
\left\vert B^{\prime}\right\vert =\sum_{B\in\pi}\left\vert B\right\vert
\left\vert U-B\right\vert $. Thus the logical information or entropy in a
partition as the normalized size of the dit set can be developed as follows:

\begin{center}
\fbox{$h\left(  \pi\right)  =$ $\frac{\sum_{B\not =B^{\prime}}\left\vert
B\right\vert \left\vert B^{\prime}\right\vert }{\left\vert U\right\vert
\times\left\vert U\right\vert }=%
{\displaystyle\sum\limits_{B\not =B^{\prime}}}
p_{B}p_{B^{\prime}}=%
{\displaystyle\sum\limits_{B\in\pi}}
p_{B}\left(  1-p_{B}\right)  =1-%
{\displaystyle\sum\limits_{B\in\pi}}
p_{B}^{2}$.}
\end{center}

Having defined and interpreted logical entropy in terms of the distinctions of
a set partition, we may, if desired, \textquotedblleft kick away the
ladder\textquotedblright\ and define the logical entropy of any finite
probability distribution $p=\{p_{1},...,p_{n}\}$ as:

\begin{center}
\fbox{$h\left(  p\right)  =\sum_{i=1}^{n}p_{i}\left(  1-p_{i}\right)
=1-\sum_{i=1}^{n}p_{i}^{2}$}.
\end{center}

\noindent The probabilistic interpretation is that $h\left(  p\right)  $ is
the probability that two independent draws (from the sample space of $n$
points with these probabilities) will give distinct points.\footnote{Note that
we can always rephrase in terms of partitions by taking $h\left(  p\right)  $
as the entropy $h\left(  \widehat{1}\right)  $ of discrete partition on
$U=\left\{  u_{1},...,u_{n}\right\}  $ with the $p_{i}$'s as the probabilities
of the singleton blocks $\left\{  u_{i}\right\}  $ of the discrete partition.}

\subsection{A Brief History of the Logical Entropy Formula: $h\left(
p\right)  =1-\sum_{i}p_{i}^{2}$}

The logical entropy formula $h\left(  p\right)  =1-\sum_{i}p_{i}^{2}$ was
motivated as the normalized count of the distinctions made by a partition,
$\left\vert \operatorname*{dit}\left(  \pi\right)  \right\vert /\left\vert
U\right\vert ^{2}$, when the probabilities are the block probabilities
$p_{B}=\frac{\left\vert B\right\vert }{\left\vert U\right\vert }$ of a
partition on a set $U$ (under a Laplacian assumption). The complementary
measure $1-h\left(  p\right)  =\sum_{i}p_{i}^{2}$ would be motivated as the
normalized count of the identifications made by a partition, $\left\vert
\operatorname*{indit}\left(  \pi\right)  \right\vert /\left\vert U\right\vert
^{2}$, thought of as an equivalence relation. Thus $1-\sum_{i}p_{i}^{2}$,
motivated by distinctions, is a measure of heterogeneity or diversity, while
the complementary measure $\sum_{i}p_{i}^{2}$, motivated by identifications,
is a measure of homogeneity or concentration. Historically, the formula can be
found in either form depending on the particular context. The $p_{i}$'s might
be relative shares such as the relative share of organisms of the $i^{th}$
species in some population of organisms, and then the interpretation of
$p_{i}$ as a probability arises by considering the random choice of an
organism from the population.

According to I. J. Good, the formula has a certain naturalness:
\textquotedblleft If $p_{1},...,p_{t}$ are the probabilities of $t$ mutually
exclusive and exhaustive events, any statistician of this century who wanted a
measure of homogeneity would have take about two seconds to suggest $\sum
p_{i}^{2}$ which I shall call $\rho$.\textquotedblright\ \cite[p.
561]{good:div} As noted by Bhargava and Uppuluri \cite{bhar:gini}, the formula
$1-\sum p_{i}^{2}$ was used by Gini in 1912 (\cite{gini:vem} reprinted in
\cite[p. 369]{gini:vemrpt}) as a measure of \textquotedblleft
mutability\textquotedblright\ or diversity. But another development of the
formula (in the complementary form) in the early twentieth century was in
cryptography. The American cryptologist, William F. Friedman, devoted a 1922
book (\cite{fried:ioc}) to the \textquotedblleft index of
coincidence\textquotedblright\ (i.e., $\sum p_{i}^{2}$). Solomon Kullback (see
the Kullback-Leibler divergence treated later) worked as an assistant to
Friedman and wrote a book on cryptology which used the index.
\cite{kull:crypt}

During World War II, Alan M. Turing worked for a time in the Government Code
and Cypher School at the Bletchley Park facility in England. Probably unaware
of the earlier work, Turing used $\rho=\sum p_{i}^{2}$ in his cryptoanalysis
work and called it the \textit{repeat rate} since it is the probability of a
repeat in a pair of independent draws from a population with those
probabilities (i.e., the identification probability $1-h\left(  p\right)  $).
Polish cryptoanalyists had independently used the repeat rate in their work on
the Enigma \cite{rej:polish}.

After the war, Edward H. Simpson, a British statistician, proposed $\sum
_{B\in\pi}p_{B}^{2}$ as a measure of species concentration (the opposite of
diversity) where $\pi$ is the partition of animals or plants according to
species and where each animal or plant is considered as equiprobable. And
Simpson gave the interpretation of this homogeneity measure as
\textquotedblleft the probability that two individuals chosen at random and
independently from the population will be found to belong to the same
group.\textquotedblright\cite[p. 688]{simp:md} Hence $1-\sum_{B\in\pi}%
p_{B}^{2}$ is the probability that a random ordered pair will belong to
different species, i.e., will be distinguished by the species partition. In
the biodiversity literature \cite{ric:unify}, the formula is known as
\textquotedblleft Simpson's index of diversity\textquotedblright\ or
sometimes, the \textquotedblleft Gini-Simpson diversity
index.\textquotedblright\ However, Simpson along with I. J. Good worked at
Bletchley during WWII, and, according to Good, \textquotedblleft E. H. Simpson
and I both obtained the notion [the repeat rate] from
Turing.\textquotedblright\ \cite[p. 395]{good:turing} When Simpson published
the index in 1948, he (again, according to Good) did not acknowledge Turing
\textquotedblleft fearing that to acknowledge him would be regarded as a
breach of security.\textquotedblright\ \cite[p. 562]{good:div}

In 1945, Albert O. Hirschman (\cite[p. 159]{hirsch:np} and \cite{hirsch:pat})
suggested using $\sqrt{\sum p_{i}^{2}}$ as an index of trade concentration
(where $p_{i}$ is the relative share of trade in a certain commodity or with a
certain partner). A few years later, Orris Herfindahl \cite{her:conc}
independently suggested using $\sum p_{i}^{2}$ as an index of industrial
concentration (where $p_{i}$ is the relative share of the $i^{th}$ firm in an
industry). In the industrial economics literature, the index $H=\sum p_{i}%
^{2}$ is variously called the Hirschman-Herfindahl index, the HH index, or
just the H index of concentration. If all the relative shares were equal
(i.e., $p_{i}=1/n$), then the identification or repeat probability is just the
probability of drawing any element, i.e., $H=1/n$, so $\frac{1}{H}=n$ is the
number of equal elements. This led to the \textquotedblleft numbers
equivalent\textquotedblright\ interpretation of the reciprocal of the H index
\cite{adel:ne}. In general, given an event with probability $p_{0}$, the
\textquotedblleft numbers-equivalent\textquotedblright\ interpretation of the
event is that it is `as if' an element was drawn out of a set of $\frac
{1}{p_{0}}$ equiprobable elements (it is `as if' since $1/p_{0}$ need not be
an integer). This numbers-equivalent idea is related to the \textquotedblleft
block-count\textquotedblright\ notion of entropy defined later.

In view of the frequent and independent discovery and rediscovery of the
formula $\rho=\sum p_{i}^{2}$ or its complement $1-\sum p_{i}^{2}$ by Gini,
Friedman, Turing, Hirschman, Herfindahl, and no doubt others, I. J. Good
wisely advises that \textquotedblleft it is unjust to associate $\rho$ with
any one person.\textquotedblright\ \cite[p. 562]{good:div}\footnote{The name
\textquotedblleft logical entropy\textquotedblright\ for $1-\sum p_{i}^{2}$
not only denotes the basic status of the formula, it avoids \textquotedblleft
Stigler's Law of Eponymy\textquotedblright: \textquotedblleft No scientific
discovery is named after its original discoverer.\textquotedblright\cite[p.
277]{stig:table}}

After Shannon's axiomatic introduction of his entropy \cite{Shannon:mtc},
there was a proliferation of axiomatic entropies with a variable
parameter.\footnote{There was no need for Shannon to present his entropy
concept axiomatically since it was based on a standard concrete interpretation
(expected number of binary partitions needed to distinguish a designated
element) which could then be generalized. The axiomatic development encouraged
the presentation of other \textquotedblleft entropies\textquotedblright\ as if
the axioms eliminated or, at least, relaxed any need for an interpretation of
the \textquotedblleft entropy\textquotedblright\ concept.} The formula $1-\sum
p_{i}^{2}$ for logical entropy appeared as a special case for a specific
parameter value in several cases. During the 1960's, Acz\'{e}l and Dar\'{o}czy
\cite{aczel:mi} developed the \textit{generalized entropies of degree }%
$\alpha$:

\begin{center}
$H_{n}^{\alpha}\left(  p_{1},...,p_{n}\right)  =\frac{\sum_{i}p_{i}^{\alpha
}-1}{\left(  2^{1-\alpha}-1\right)  }$
\end{center}

\noindent and the logical entropy occurred as half the value for $\alpha=2$.
That formula also appeared as Havrda-Charvat's \textit{structural }$\alpha
$\textit{-entropy} \cite{hav:alpha}:

\begin{center}
$S(p_{1},...,p_{n},;\alpha)=\frac{2^{\alpha-1}}{2^{\alpha-1}-1}\left(
1-\sum_{i}p_{i}^{\alpha}\right)  $
\end{center}

\noindent and the special case of $\alpha=2$ was considered by Vajda
\cite{vajda:patt}.

Patil and Taillie \cite{patil:div} defined the \textit{diversity index of
degree }$\beta$ in 1982:

\begin{center}
$\Delta_{\beta}=\frac{1-\sum_{i}p_{i}^{\beta+1}}{\beta}$
\end{center}

\noindent and Tsallis \cite{tsallis:pg} independently gave the same formula as
an entropy formula in 1988:

\begin{center}
$S_{q}(p_{1},...,p_{n})=\frac{1-\sum_{i}p_{i}^{q}}{q-1}$
\end{center}

\noindent where the logical entropy formula occurs as a special case
($\beta=1$ or $q=2$). While the generalized parametric entropies may be
interesting as axiomatic exercises, our purpose is to emphasize the specific
logical interpretation of the logical entropy formula (or its complement).

From the logical viewpoint, two elements from $U=\left\{  u_{1},...,u_{n}%
\right\}  $ are either identical or distinct. Gini \cite{gini:vem} introduced
$d_{ij}$ as the \textquotedblleft distance\textquotedblright\ between the
$i^{th}$ and $j^{th}$ elements where $d_{ij}=1$ for $i\not =j$ and $d_{ii}=0$.
Since $1=\left(  p_{1}+...+p_{n}\right)  \left(  p_{1}+...+p_{n}\right)
=\sum_{i}p_{i}^{2}+\sum_{i\not =j}p_{i}p_{j}$, the logical entropy, i.e.,
Gini's index of mutability, $h\left(  p\right)  =1-\sum_{i}p_{i}^{2}%
=\sum_{i\not =j}p_{i}p_{j}$, is the average logical distance between a pair of
independently drawn elements. But one might generalize by allowing other
distances $d_{ij}=d_{ji}$ for $i\not =j$ (but always $d_{ii}=0$) so that
$Q=\sum_{i\not =j}d_{ij}p_{i}p_{j}$ would be the average distance between a
pair of independently drawn elements from $U$. In 1982, C. R. (Calyampudi
Radhakrishna) Rao introduced precisely this concept as \textit{quadratic
entropy} \cite{rao:div} (which was later rediscovered in the biodiversity
literature as the \textquotedblleft Avalanche Index\textquotedblright\ by
Ganeshaish et al. \cite{gane:aval}). In many domains, it is quite reasonable
to move beyond the bare-bones logical distance of $d_{ij}=1$ for $i\not =j$ so
that Rao's quadratic entropy is a useful and easily interpreted generalization
of logical entropy.

\section{Relationship between the Logical and Shannon Entropies}

\subsection{The Search Approach to Find the \textquotedblleft Sent
Message\textquotedblright}

The logical entropy $h\left(  \pi\right)  =\sum_{B\in\pi}p_{B}\left(
1-p_{B}\right)  $ in this form as an average over blocks allows a direct
comparison with Shannon's entropy $H\left(  \pi\right)  =\sum_{B\in\pi}%
p_{B}\log_{2}(\frac{1}{p_{B}})$ of the partition which is also an average over
the blocks. What is the connection between the \textit{block entropies}
$h\left(  B\right)  =1-p_{B}$ and $H\left(  B\right)  =\log_{2}\left(
\frac{1}{p_{B}}\right)  $? Shannon uses reasoning (shared with Hartley) to
arrive at a notion of entropy or information content for an element out of a
subset (e.g., a block in a partition as a set of blocks). Then for a partition
$\pi$, Shannon averages the block values to get the partition value $H\left(
\pi\right)  $. Hartley and Shannon start with the question of the information
required to single an element $u$ out of a set $U$, e.g., to single out the
sent message from the set of possible messages. Alfred Renyi has also
emphasized this \textquotedblleft search-theoretic\textquotedblright\ approach
to information theory (see \cite{Renyi:trs}, \cite{Renyi:pt}, or numerous
papers in \cite{Renyi:sp}).\footnote{In Gian-Carlo Rota's teaching, he
supposed that the Devil had picked an element out of $U$ and would not reveal
its identity. But when given a binary partition (i.e., a yes-or-no question),
the Devil had to truthfully tell which block contained the hidden element.
Hence the problem was to find the minimum number of binary partitions needed
to force the Devil to reveal the hidden element.}

One intuitive measure of the information obtained by determining the
designated element in a set $U$ of equiprobable elements would just be the
cardinality $\left\vert U\right\vert $ of the set, and, as we will see, that
leads to a multiplicative \textquotedblleft block-count\textquotedblright%
\ version of Shannon's entropy. But Hartley and Shannon wanted the additivity
that comes from taking the logarithm of the set size $\left\vert U\right\vert
$. If $\left\vert U\right\vert =2^{n}$ then this allows the crucial Shannon
interpretation of $\log_{2}\left(  \left\vert U\right\vert \right)  =n$ as the
minimum number of yes-or-no questions (binary partitions) it takes to single
out any designated element (the \textquotedblleft sent
message\textquotedblright) of the set. In a mathematical version of the game
of twenty questions (like R\'{e}nyi's Hungarian game of \textquotedblleft
Bar-Kochba\textquotedblright), think of each element of $U$ as being assigned
a unique binary number with $n$ digits. Then the minimum $n$ questions can
just be the questions asking for the $i^{th}$ binary digit of the hidden
designated element. Each answer gives one \textit{bit} (short for
\textquotedblleft binary digit\textquotedblright) of information. With this
motivation for the case of $\left\vert U\right\vert =2^{n}$, Shannon and
Hartley take $\log\left(  \left\vert U\right\vert \right)  $ as the measure of
the information required to single out a hidden element in a set with
$\left\vert U\right\vert $ equiprobable elements.\footnote{Hartley used logs
to the base $10$ but here all logs are to base $2$ unless otherwise indicated.
Instead of considering whether the base should be $2$, $10$, or $e$, it is
perhaps more important to see that there is a natural base-free variation
$H_{m}\left(  \pi\right)  $ \ on Shannon's entropy (see \textquotedblleft
block-count entropy\textquotedblright\ defined below).} That extends the
\textquotedblleft minimum number of yes-or-no questions\textquotedblright%
\ motivation from $\left\vert U\right\vert =2^{n}$ to any finite\ set $U$ with
$\left\vert U\right\vert $ equiprobable elements. If a partition $\pi$ had
equiprobable blocks, then the Shannon entropy would be $H\left(  B\right)
=\log\left(  \left\vert \pi\right\vert \right)  $ where $\left\vert
\pi\right\vert $ is the number of blocks.

To extend this basic idea to sets of elements which are not equiprobable
(e.g., partitions with unequal blocks), it is useful to use an old device to
restate any positive probability as a chance among equiprobable elements. If
$p_{i}=0.02$, then there is a $1$ in $50=\frac{1}{p_{i}}$ chance of the
$i^{th}$ outcome occurring in any trial. It is \textquotedblleft as
if\textquotedblright\ the outcome was one among $1/p_{i}$ equiprobable
outcomes.\footnote{Since $1/p_{i}$ need not be an integer (or even rational),
one could interpret the equiprobable \textquotedblleft number of
elements\textquotedblright\ as being heuristic or one could restate it in
continuous terms. The continuous version is the uniform distribution on the
real interval $\left[  0,1/p_{i}\right]  $ where the probability of an outcome
in the unit interval $\left[  0,1\right]  $ is $1/\left(  1/p_{i}\right)
=p_{i}$.} Thus each positive probability $p_{i}$ has an associated
\textit{equivalent number} $1/p_{i}$ which is the size of the hypothetical set
of equiprobable elements so that the probability of drawing any given element
is $p_{i}$.\footnote{In continuous terms, the \textit{numbers-equivalent} is
the length of the interval $\left[  0,1/p_{i}\right]  $ with the uniform
distribution on it.}

Given a partition $\left\{  B\right\}  _{B\in\pi}$ with unequal blocks, we
motivate the block entropy $H\left(  B\right)  $ for a block with probability
$p_{B}$ by taking it as the entropy for a hypothetical
\textit{numbers-equivalent partition} $\pi_{B}$ with $\frac{1}{p_{B}}$
equiprobable blocks, i.e.,

\begin{center}
$H\left(  B\right)  =\log\left(  \left\vert \pi_{B}\right\vert \right)
=\log\left(  \frac{1}{p_{B}}\right)  $.
\end{center}

\noindent With this motivation, the Shannon entropy of the partition is then
defined as the arithmetical average of the block entropies:

\begin{center}
Shannon's entropy: $H\left(  \pi\right)  =%
{\displaystyle\sum\limits_{B\in\pi}}
p_{B}H(B)=%
{\displaystyle\sum\limits_{B\in\pi}}
p_{B}\log\left(  \frac{1}{p_{B}}\right)  $.
\end{center}

This can be directly compared to the logical entropy $h\left(  \pi\right)
=\sum_{B\in\pi}p_{B}\left(  1-p_{B}\right)  $ which arose from quite different
distinction-based reasoning (e.g., where the search of a single designated
element played no role). Nevertheless, the formula $\sum_{B\in\pi}p_{B}\left(
1-p_{B}\right)  $ can be viewed as an average over the quantities which play
the role of \textquotedblleft block entropies\textquotedblright\ $h\left(
B\right)  =\left(  1-p_{B}\right)  $. But this \textquotedblleft block
entropy\textquotedblright\ cannot be directly interpreted as a (normalized)
dit count since there is no such thing as the dit count for a single block.
The dits are the pairs of elements in \textit{distinct} blocks.

For comparison purposes, we may nevertheless carry over the heuristic
reasoning to the case of logical entropy. For each block $B$, we take the same
hypothetical numbers-equivalent partition $\pi_{B}$ with $\frac{\left\vert
U\right\vert }{\left\vert B\right\vert }=\frac{1}{p_{B}}$ equal blocks of size
$\left\vert B\right\vert $ and then take the desired block entropy $h\left(
B\right)  $ as the normalized dit count $h\left(  \pi_{B}\right)  $ for that
partition. Each block contributes $p_{B}\left(  1-p_{B}\right)  $ to the
normalized dit count and there are $\left\vert U\right\vert /\left\vert
B\right\vert =1/p_{B}$ blocks in $\pi_{B}$ so the total normalized dit count
simplifies to: $h\left(  \pi_{B}\right)  =\frac{1}{p_{B}}p_{B}\left(
1-p_{B}\right)  =1-p_{B}=h\left(  B\right)  $, which we could take as the
\textit{logical block entropy}. Then the average of these logical block
entropies gives the logical entropy $h\left(  \pi\right)  =\sum_{B\in\pi}%
p_{B}h\left(  B\right)  =\sum_{B\in\pi}p_{B}\left(  1-p_{B}\right)  $ of the
partition $\pi$, all in the manner of the heuristic development of Shannon's
$H\left(  \pi\right)  =\sum_{B\in\pi}p_{B}\log\left(  \frac{1}{p_{B}}\right)
$.

There is, however, no need to go through this reasoning to arrive at the
logical entropy of a partition as the average of block entropies. The
interpretation of the logical entropy as the normalized dit count survives the
averaging even though all the blocks of $\pi$ might have different sizes,
i.e., the interpretation \textquotedblleft commutes\textquotedblright\ with
the averaging of block entropies. Thus $h\left(  \pi\right)  $ is the
\textit{actual} dit count (normalized) for a partition $\pi$, not just the
average of block entropies $h\left(  B\right)  $ that could be interpreted as
the normalized dit counts for hypothetical partitions $\pi_{B}$.

The interpretation of the Shannon measure of information as the minimum number
of binary questions it takes to single out a designated block does not commute
with the averaging over the set of different-sized blocks in a partition.
Hence the Shannon entropy of a partition is the \textit{expected} number of
bits it takes to single out the designated block while the logical entropy of
a partition on a set is the \textit{actual} number of dits (normalized)
distinguished by the partition.

The last step in connecting Shannon entropy and logical entropy is to rephrase
the heuristics behind Shannon entropy in terms of \textquotedblleft making all
the distinctions\textquotedblright\ rather than \textquotedblleft singling out
the designated element.\textquotedblright

\subsection{Distinction-based Treatment of Shannon's Entropy}

The search-theoretic approach was the heritage of the original application of
information theory to communications where the focus was on singling out a
designated element, the sent message. In the \textquotedblleft twenty
questions\textquotedblright\ version, one person picks a hidden element and
the other person seeks the minimum number of binary partitions on the set of
possible answers to single out the answer. But it is simple to see that the
focus on the single designated element was unnecessary. The essential point
was \textit{to make all the distinctions to separate the elements}--since any
element could have been the designated one. If the join of the minimum number
of binary partitions did not distinguish all the elements into singleton
blocks, then one could not have picked out the hidden element if it was in a
non-singleton block. Hence the distinction-based treatment of Shannon's
entropy amounts to rephrasing the above heuristic argument in terms of
\textquotedblleft making all the distinctions\textquotedblright\ rather than
\textquotedblleft making the distinctions necessary to single out any
designated element.\textquotedblright\ 

In the basic example of $\left\vert U\right\vert =2^{n}$ where we may think of
the $2^{n}$ like or equiprobable elements as being encoded with $n$ binary
digit numbers, then $n=\log\left(  \frac{1}{1/2^{n}}\right)  $ is the minimum
number of binary partitions (each partitioning according to one of the $n$
digits) necessary \textit{to make all the distinctions} between the elements,
i.e., the minimum number of binary partitions whose join is the discrete
partition with singleton blocks (each block probability being $p_{B}=1/2^{n}%
$). Generalizing to any set $U$ of equiprobable elements, the minimum number
of bits necessary to distinguish all the elements from each other is
$\log\left(  \frac{1}{1/\left\vert U\right\vert }\right)  =\log\left(
\left\vert U\right\vert \right)  $. Given a partition $\pi=\left\{  B\right\}
_{B\in\pi}$ on $U$, the block entropy $H\left(  B\right)  =\log\left(
\frac{1}{p_{B}}\right)  $ is the minimum number of bits necessary to
distinguish all the blocks in the numbers-equivalent partition $\pi_{B}$, and
the average of those block entropies gives the Shannon entropy: $H\left(
\pi\right)  =\sum_{B\in\pi}p_{B}\log\left(  \frac{1}{p_{B}}\right)  $.

The point of rephrasing the heuristics behind Shannon's definition of entropy
in terms of the average bits needed to \textquotedblleft make all the
distinctions\textquotedblright\ is that it can then be directly compared with
the logical definition of entropy which is simply the total number of
distinctions normalized by $\left\vert U\right\vert ^{2}$. Thus the two
definitions of entropy boil down to two different ways of measuring the
totality of distinctions. A third way to measure the totality of distinctions,
called the \textquotedblleft block-count entropy,\textquotedblright\ is
defined below. Hence we have our overall theme that these three notions of
entropy boil down to three ways of \textquotedblleft counting
distinctions.\textquotedblright

\subsection{Relationships Between the Block Entropies}

Since the logical and Shannon entropies have formulas presenting them as
averages of block-entropies, $h\left(  \pi\right)  =\sum_{B\in\pi}p_{B}\left(
1-p_{B}\right)  $ and $H\left(  \pi\right)  =\sum_{B\in\pi}p_{B}\log\left(
\frac{1}{p_{B}}\right)  $, the two notions are precisely related by their
respective block entropies, $h\left(  B\right)  =1-p_{B}$ and $H\left(
B\right)  =\log\left(  \frac{1}{p_{B}}\right)  $. Solving each for $p_{B}$ and
then eliminating it yields the:

\begin{center}
\fbox{Block entropy relationship: $h\left(  B\right)  =1-\frac{1}{2^{H\left(
B\right)  }}$ and $H\left(  B\right)  =\log\left(  \frac{1}{1-h\left(
B\right)  }\right)  $}.
\end{center}

The block entropy relation, $h\left(  B\right)  =1-\frac{1}{2^{H\left(
B\right)  }}$, has a simple probabilistic interpretation. Thinking of
$H\left(  B\right)  $ as an integer, $H\left(  B\right)  $ is the Shannon
entropy of the discrete partition on $U$ with $\left\vert U\right\vert
=2^{H\left(  B\right)  }$ elements while $h\left(  B\right)  =1-\frac
{1}{2^{H\left(  B\right)  }}=1-p_{B}$ is the logical entropy of that partition
since $1/2^{H\left(  B\right)  }$ is the probability of each block in that
discrete partition. The probability that a random pair is distinguished by a
discrete partition is just the probability that the second draw is distinct
from the first draw. Given the first draw from a set of $2^{H\left(  B\right)
}$ individuals, the probability that the second draw (with replacement) is
different is $1-\frac{1}{2^{H\left(  B\right)  }}=h\left(  B\right)  $.

To summarize the comparison up to this point, the logical theory and Shannon's
theory start by posing different questions which then turn out to be precisely
related. Shannon's statistical theory of communications is concerned with
determining the sent message out of a set of possible messages. In the basic
case, the messages are equiprobable so it is abstractly the problem of
determining the hidden designated element out of a set of equiprobable
elements which, for simplicity, we can assume has $2^{n}$ elements. The
process of determining the hidden element can be conceptualized as the process
of asking binary questions which split the set of possibilities into
equiprobable parts. The answer to the first question determines which subset
of $2^{n-1}$ elements contains the hidden element and that provides $1$ bit of
information. An independent equal-blocked binary partition would split each of
the $2^{n-1}$ element blocks into equal blocks with $2^{n-2}$ elements each.
Thus $2$ bits of information would determine which of those $2^{2}$ blocks
contained the hidden element, and so forth. Thus $n$ independent equal-blocked
binary partitions would determine which of the resulting $2^{n}$ blocks
contains the hidden element. Since there are $2^{n}$ elements, each of those
blocks is a singleton so the hidden element has been determined. Hence the
problem of finding a designated element among $2^{n}$ equiprobable elements
requires $\log\left(  2^{n}\right)  =n$ bits of information.

The logical theory starts with the basic notion of a distinction between
elements and defines the logical information in a set of distinct $2^{n}$
elements as the (normalized) number of distinctions that need to be made to
distinguish the $2^{n}$ elements. The distinctions are counted as ordered
rather than unordered pairs (in order to better apply the machinery of
probability theory) and the number of distinctions or dits is normalized by
the number of all ordered pairs. Hence a set of $2^{n}$ distinct elements
would involve $\left\vert U\times U-\Delta_{U}\right\vert =2^{n}\times
2^{n}-2^{n}=2^{2n}-2^{n}=2^{n}\left(  2^{n}-1\right)  $ distinctions which
normalizes to $\frac{2^{2n}-2^{n}}{2^{2n}}=1-\frac{1}{2^{n}}$.

There is, however, no need to motivate Shannon's entropy by focusing on the
search for a designated element. The task can equivalently be taken as
distinguishing all elements from each other rather than distinguishing a
designated element from all the other elements. The connection between the two
approaches can be seen by computing the total number of distinctions made by
intersecting the $n$ independent equal-blocked binary partitions in Shannon's approach.

\begin{description}
\item[Example of counting distinctions:] Doing the computation, the first
partition which creates two sets of $2^{n-1}$ elements each thereby creates
$2^{n-1}\times2^{n-1}=2^{2n-2}$ distinctions as unordered pairs and
$2\times2^{2n-2}=2^{2n-1}$ distinctions as ordered pairs. The next binary
partition splits each of those blocks into equal blocks of $2^{n-2}$ elements.
Each split block creates $2^{n-2}\times2^{n-2}=2^{2n-4}$ new distinctions as
unordered pairs and there were two such splits so there are $2\times
2^{2n-4}=2^{2n-3}$ additional unordered pairs of distinct elements created or
$2^{2n-2}$ new ordered pair distinctions. In a similar manner, the third
partition creates $2^{2n-3}$ new dits and so forth down to the $n^{th}$
partition which adds $2^{2n-n}$ new dits. Thus in total, the intersection of
the $n$ independent equal-blocked binary partitions has created $2^{2n-1}%
+2^{2n-2}+...+2^{2n-n}=2^{n}\left(  2^{n-1}+2^{n-2}+...+2^{0}\right)
=2^{n}\left(  \frac{2^{n}-1}{2-1}\right)  =2^{n}\left(  2^{n}-1\right)  $
(ordered pair) distinctions which are all the dits on a set with $2^{n}$
elements. This is the instance of the block entropy relationship $h\left(
B\right)  =1-\frac{1}{2^{H\left(  B\right)  }}$ when the block $B$ is a
singleton in a $2^{n}$ element set so that $H\left(  B\right)  =\log\left(
\frac{1}{1/2^{n}}\right)  =\log\left(  2^{n}\right)  =n$ and $h\left(
B\right)  =1-\frac{1}{2^{H\left(  B\right)  }}=1-\frac{1}{2^{n}}$.
\end{description}

Thus the Shannon entropy as the number of independent equal-blocked binary
partitions it takes to single out a hidden designated element in a $2^{n}$
element set is \textit{also} the number of independent equal-blocked binary
partitions it takes to distinguish all the elements of a $2^{n}$ element set
from each other.

The connection between Shannon entropy and logical entropy boils down to two points.

\begin{enumerate}
\item The first point is the basic fact that for binary partitions to single
out a hidden element (\textquotedblleft sent message\textquotedblright) in a
set is the same as the partitions distinguishing any pair of distinct elements
(since if a pair was left undistinguished, the hidden element could not be
singled out if it were one of the elements in that undistinguished pair). This
gives what might be called the \textit{distinction interpretation} of Shannon
entropy as a count of the binary partitions necessary to distinguish between
all the distinct messages in the set of possible messages in contrast to the
usual \textit{search interpretation} as the binary partition count necessary
to find the hidden designated element such as the sent message.

\item The second point is that in addition to the Shannon count of the binary
partitions necessary to make all the distinctions, we may use the logical
measure that is simply the (normalized) count of the distinctions themselves.
\end{enumerate}

\subsection{A Coin-Weighing Example}

The logic of the connection between joining independent equal-blocked
partitions and efficiently creating dits is not dependent on the choice of
base $2$. Consider the coin-weighing problem where one has a balance scale and
a set of $3^{n}$ coins all of which look alike but one is counterfeit (the
hidden designated element) and is lighter than the others. The coins might be
numbered using the $n$-digit numbers in mod $3$ arithmetic where the three
digits are $0$, $1$, and $2$. The $n$ independent ternary partitions are
arrived at by dividing the coins into three piles according to the $i^{th}$
digit as $i=1,...,n$. To use the $n$ partitions to find the false coin, two of
the piles are put on the balance scale. If one side is lighter, then the
counterfeit coin is in that block. If the two sides balance, then the light
coin is in the third block of coins not on the scale. Thus $n$ weighings
(i.e., the join of $n$ independent equal-blocked ternary partitions) will
determine the $n$ ternary digits of the false coin, and thus the ternary
Shannon entropy is $\log_{3}\left(  \frac{1}{1/3^{n}}\right)  =\log_{3}\left(
3^{n}\right)  =n$ trits. As before we can interpret the joining of independent
partitions not only as the most efficient way to find the hidden element
(e.g., the false coin or the sent message) but as the most efficient way to
make all the distinctions between the elements of the set.

The first partition (separating by the first ternary digit) creates $3$ equal
blocks of $3^{n-1}$ elements each so that creates $3\times3^{n-1}\times
3^{n-1}=3^{2n-1}$ unordered pairs of distinct elements or $2\times3^{2n-1}$
ordered pair distinctions. The partition according to the second ternary digit
divides each of these three blocks into three equal blocks of $3^{n-2}$
elements each so the additional unordered pairs created are $3\times
3\times3^{n-2}\times3^{n-2}=3^{2n-2}$ or $2\times3^{2n-2}$ ordered pair
distinctions. Continuing in this fashion, the $n^{th}$ ternary partition adds
$2\times3^{2n-n}$ dits. Hence the total number of dits created by joining the
$n$ independent partitions is:

\begin{center}
$2\times\left[  3^{2n-1}+3^{2n-2}...+3^{n}\right]  =2\times\left[
3^{n}\left(  3^{n-1}+3^{n-2}...+1\right)  \right]  =2\times\left[  3^{n}%
\frac{(3^{n}-1)}{3-1}\right]  =3^{n}\left(  3^{n}-1\right)  $
\end{center}

\noindent which is the total number of ordered pair distinctions between the
elements of the $3^{n}$ element set. Thus the Shannon measure in trits is the
minimum number of ternary partitions needed to create all the distinctions
between the elements of a set. The base-$3$ Shannon entropy is $H_{3}\left(
\pi\right)  =\sum_{B\in\pi}p_{B}\log_{3}\left(  \frac{1}{p_{B}}\right)  $
which for this example of the discrete partition on a $3^{n}$ element set $U$
is $H_{3}\left(  \widehat{1}\right)  =\sum_{u\in U}\frac{1}{3^{n}}\log
_{3}\left(  \frac{1}{1/3^{n}}\right)  =\log_{3}\left(  3^{n}\right)  =n$ which
can also be thought of as the block value entropy for a singleton block so
that we may apply the block value relationship. The logical entropy of the
discrete partition on this set is: $h\left(  \widehat{1}\right)  =\frac
{3^{n}\left(  3^{n}-1\right)  }{3^{2n}}=1-\frac{1}{3^{n}}$ which could also be
thought of as the block value of the logical entropy for a singleton block.
Thus the entropies for the discrete partition stand in the block value
relationship which for base $3$ is:

\begin{center}
$h\left(  B\right)  =1-\frac{1}{3^{H_{3}\left(  B\right)  }}$.
\end{center}

The example helps to show how the logical notion of a distinction underlies
the Shannon measure of information, and how a complete procedure for finding
the hidden element (e.g., the sent message) is equivalent to being able to
make all the distinctions in a set of elements. But this should not be
interpreted as showing that the Shannon's information theory \textquotedblleft
reduces\textquotedblright\ to the logical theory. The Shannon theory is
addressing an additional question of finding the unknown element. One can have
all the distinctions between elements, e.g., the assignment of distinct
base-$3$ numbers to the $3^{n}$ coins, without knowing which element is the
designated one. Information theory becomes a theory of the
\textit{transmission} of information, i.e., a theory of communication, when
that second question of \textquotedblleft receiving the
message\textquotedblright\ as to which element is the designated one is the
focus of analysis. In the coin example, we might say that the information
about the light coin was always there in the nature of the situation (i.e.,
taking \textquotedblleft nature\textquotedblright\ as the sender) but was
unknown to an observer (i.e., on the receiver side). The coin weighing scheme
was a way for the observer to elicit the information out of the situation.
Similarly, the game of twenty questions is about finding a way to uncover the
hidden answer---which was all along distinct from the other possible answers
(on the sender side). It is this question of the transmission of information
(and the noise that might interfere with the process) that carries Shannon's
statistical theory of communications well beyond the bare-bones logical
analysis of information in terms of distinctions.

\subsection{Block-count Entropy}

The fact that the Shannon motivation works for other bases than $2$ suggests
that there might be a base-free version of the Shannon measure (the logical
measure is already base-free). Sometimes the reciprocal $\frac{1}{p_{B}}$ of
the probability of an event $B$ is interpreted as the \textquotedblleft
surprise-value information\textquotedblright\ conveyed by the occurrence of
$B$. But there is a better concept to use than the vague notion of
\textquotedblleft surprise-value information.\textquotedblright\ For any
positive probability $p_{0}$, we defined the reciprocal $\frac{1}{p_{0}}$ as
the\textit{\ equivalent number }of (equiprobable) elements (always
\textquotedblleft as it were\textquotedblright\ since it need not be an
integer) since that is the number of equiprobable elements in a set so that
the probability of choosing any particular element is $p_{0}$. The
\textquotedblleft big surprise\textquotedblright\ as a small probability event
occurs means it is \textquotedblleft as if\textquotedblright\ a particular
element was picked from a big set of elements. For instance, for a block
probability $p_{B}=\frac{\left\vert B\right\vert }{\left\vert U\right\vert }$,
its numbers-equivalent is the number of blocks $\frac{\left\vert U\right\vert
}{\left\vert B\right\vert }=\frac{1}{p_{B}}$ in the hypothetical equal-blocked
partition $\pi_{B}$ with each block equiprobable with $B$. Our task is to
develop this number-of-blocks\ or block-count measure of information for partitions.

The \textit{block-count block entropy} $H_{m}\left(  B\right)  $ is just the
number of blocks in the hypothetical number-of-equivalent-blocks partition
$\pi_{B}$ where $B$ is one of $\frac{\left\vert U\right\vert }{\left\vert
B\right\vert }=\frac{1}{p_{B}}$ associated similar blocks so that
$H_{m}\left(  B\right)  =\frac{1}{p_{B}}$.

If events $B$ and $C$ were independent, then $p_{B\cap C}=p_{B}p_{C}$ so the
equivalent number of elements associated with the occurrence of both events is
the product $\frac{1}{p_{B\cap C}}=\frac{1}{p_{B}}\frac{1}{p_{C}}$ of the
number of elements associated with the separate events. This suggests that the
average of the block entropies $H_{m}\left(  B\right)  =\frac{1}{p_{B}}$
should be the multiplicative average (or geometric mean) rather than the
arithmetical average.

Hence we define the \textit{number-of-equivalent blocks entropy }or, in short,
\textit{block-count entropy} of a partition $\pi$ (which does not involve any
choice of a base for logs) as the geometric mean of block entropies:

\begin{center}
\fbox{Block-count entropy: $H_{m}\left(  \pi\right)  =%
{\textstyle\prod\limits_{B\in\pi}}
H_{m}\left(  B\right)  ^{p_{B}}=%
{\textstyle\prod\limits_{B\in\pi}}
\left(  \frac{1}{p_{B}}\right)  ^{p_{B}}$ blocks}.
\end{center}

\noindent Finding the designated block in $\pi$ is the same on average\ as
finding the designated block in a partition with $H_{m}\left(  \pi\right)  $
equal blocks. But since $H_{m}\left(  \pi\right)  $ need not be an integer,
one might take the reciprocal to obtain the probability interpretation:
finding the designated block in $\pi$ is the same on average\ as the
occurrence of an event with probability $1/H_{m}\left(  \pi\right)  $.

Given a finite-valued random variable $X$ with the values $\left\{
x_{1},...,x_{n}\right\}  $ with the probabilities $\left\{  p_{1}%
,...,p_{n}\right\}  $, the \textit{additive expectation} is: $E\left[
X\right]  =\sum_{i=1}^{n}p_{i}x_{i}$ and the \textit{multiplicative
expectation} is: $E_{m}\left[  X\right]  =\prod_{i=1}^{n}x_{i}^{p_{i}}$.
Treating the block probability as a random variable defined on the blocks of a
partition, all three entropies can be expressed as expectations:

\begin{center}%
\begin{align*}
H_{m}\left(  \pi\right)   &  =E_{m}\left[  \frac{1}{p_{B}}\right] \\
H\left(  \pi\right)   &  =E\left[  \log\left(  \frac{1}{p_{B}}\right)  \right]
\\
h\left(  \pi\right)   &  =E\left[  1-p_{B}\right]  =1-E\left[  p_{B}\right]  .
\end{align*}

\end{center}

The usual (additive) Shannon entropy is then obtained as the $\log_{2}$
version of this \textquotedblleft log-free\textquotedblright\ block-count entropy:

\begin{center}
$\log_{2}\left(  H_{m}\left(  \pi\right)  \right)  =\log\left(
{\textstyle\prod\limits_{B\in\pi}}
\left(  \frac{1}{p_{B}}\right)  ^{p_{B}}\right)  =\sum_{B\in\pi}\log\left(
\left(  \frac{1}{p_{B}}\right)  ^{p_{B}}\right)  =\sum_{B\in\pi}p_{B}%
\log\left(  \frac{1}{p_{B}}\right)  =H\left(  \pi\right)  $.
\end{center}

\noindent Or viewed the other way around, $H_{m}\left(  \pi\right)
=2^{H\left(  \pi\right)  }$.\footnote{Thus we expect the number-of-blocks
entropy to be multiplicative where the usual Shannon entropy is additive
(e.g., for stochastically independent partitions) and hence the subscript on
$H_{m}\left(  \pi\right)  $.} The base $3$ entropy encountered in the
coin-weighing example is obtained by taking logs to that base: $H_{3}\left(
\pi\right)  =\log_{3}\left(  H_{m}\left(  \pi\right)  \right)  $, and
similarly for the Shannon entropy with natural logs: $H_{e}\left(  \pi\right)
=\log_{e}\left(  H_{m}\left(  \pi\right)  \right)  $, or with common logs:
$H_{10}\left(  \pi\right)  =\log_{10}\left(  H_{m}\left(  \pi\right)  \right)
$.

Note that this relation $H_{m}\left(  \pi\right)  =2^{H\left(  \pi\right)  }$
is a result, not a definition. The block-count entropy was defined from
\textquotedblleft scratch\textquotedblright\ in a manner similar to the usual
Shannon entropy (which thus might be called the \textquotedblleft$\log_{2}%
$-of-block-count entropy\textquotedblright\ or \textquotedblleft
binary-partition-count entropy\textquotedblright). In a partition of
individual organisms by species, the interpretation of $2^{H\left(
\pi\right)  }$ (or $e^{H_{e}\left(  \pi\right)  }$ when natural logs are used)
is the \textquotedblleft number of equally common species\textquotedblright%
\ \cite[p. 514]{maca:div}. MacArthur argued that this block-count entropy
(where a block is a species) will \textquotedblleft accord much more closely
with our intuition...\textquotedblright\ (than the usual Shannon entropy).

The block-count entropy is the information measure that takes the count of a
set (of like elements) as the measure of the information in the set. That is,
for the discrete partition on $U$, each $p_{B}$ is $\frac{1}{\left\vert
U\right\vert }$ so the block-count entropy of the discrete partition is
$H_{m}\left(  \widehat{1}\right)  =%
{\displaystyle\prod\limits_{u\in U}}
\left\vert U\right\vert ^{1/\left\vert U\right\vert }=\left\vert U\right\vert
$ which could also be obtained as $2^{H\left(  \widehat{1}\right)  }$ since
$H\left(  \widehat{1}\right)  =\log\left(  \left\vert U\right\vert \right)  $
is the $\log_{2}$-of-block-count Shannon entropy of $\widehat{1}$. Hence, the
natural choice of unit for the block-count entropy is \textquotedblleft
blocks\textquotedblright\ (as in $H_{m}\left(  \widehat{1}\right)  =\left\vert
U\right\vert $ blocks in the discrete partition on $U$). The block-count
entropy of the discrete partition on an equiprobable $3^{n}$ element set is
$3^{n}$ blocks.\ Hence the Shannon entropy with base $3$ would be the
$\log_{3}$-of-block-count entropy: $\log_{3}\left(  H_{m}\left(  \widehat
{1}\right)  \right)  =\log_{3}\left(  3^{n}\right)  =n$ trits as in the
coin-weighing example above. The block value relationship between the
block-count entropy and the logical entropy in general is:

\begin{center}
$h(B)=1-p_{B}=1-\frac{1}{1/p_{B}}=1-\frac{1}{H_{m}\left(  B\right)  }%
=1-\frac{1}{2^{H\left(  B\right)  }}=1-\frac{1}{3^{H_{3}\left(  B\right)  }%
}=1-\frac{1}{e^{H_{e}\left(  B\right)  }}=1-\frac{1}{10^{H_{10}\left(
B\right)  }}$
\end{center}

\noindent where $H_{m}\left(  B\right)  =1/p_{B}=2^{H\left(  B\right)
}=3^{H_{3}\left(  B\right)  }=e^{H_{e}\left(  B\right)  }=10^{H_{10}\left(
B\right)  }$.

\section{Analogous Concepts for Shannon and Logical Entropies}

\subsection{Independent Partitions}

It is sometimes asserted that \textquotedblleft information\textquotedblright%
\ should be additive for independent\footnote{Recall the \textquotedblleft
independent\textquotedblright\ means stochastic independence so that
partitions $\pi$ and $\sigma$ are \textit{independent} if for all $B\in\pi$
and $C\in\sigma$, $p_{B\cap C}=p_{B}p_{C}$.} partitions but the underlying
mathematical fact is that the block-count is multiplicative for independent
partitions and Shannon chose to use the logarithm of the block-count as his
measure of information.

If two partitions $\pi=\left\{  B\right\}  _{B\in\pi}$ and $\sigma=\left\{
C\right\}  _{C\in\sigma}$ are independent, then the block\ counts (i.e., the
block entropies for the block-count entropy) multiply, i.e., $H_{m}(B\cap
C)=\frac{1}{p_{B\cap C}}=\frac{1}{p_{B}}\frac{1}{p_{C}}=H_{m}\left(  B\right)
H_{m}\left(  C\right)  $. Hence for the multiplicative expectations we have:

\begin{center}
$H_{m}(\pi\vee\sigma)=\prod_{B,C}H_{m}\left(  B\cap C\right)  ^{p_{B\cap C}%
}=\prod_{B,C}\left[  H_{m}(B)H_{m}\left(  C\right)  \right]  ^{p_{B}p_{C}%
}=\left(  \prod_{B\in\pi}H_{m}\left(  B\right)  ^{p_{B}}\right)  \left(
\prod_{C\in\sigma}H_{m}(C)^{p_{C}}\right)  =H_{m}\left(  \pi\right)
H_{m}\left(  \sigma\right)  $,
\end{center}

\noindent or taking logs to any desired base such as $2$:

\begin{center}
$H\left(  \pi\vee\sigma\right)  =\log_{2}(H_{m}(\pi\vee\sigma))=\log
_{2}\left(  H_{m}\left(  \pi\right)  H_{m}\left(  \sigma\right)  \right)
=\log_{2}\left(  H_{m}\left(  \pi\right)  \right)  +\log_{2}\left(
H_{m}\left(  \sigma\right)  \right)  =H\left(  \pi\right)  +H\left(
\sigma\right)  $.
\end{center}

Thus for independent partitions, the block-count entropies multiply and the
log-of-block-count entropies add. What happens to the logical entropies? We
have seen that when the information in a partition is represented by its dit
set $\operatorname*{dit}\left(  \pi\right)  $, then the overlap in the dit
sets of any two non-blob partitions is always non-empty. The dit set of the
join of two partitions is just the union, $\operatorname*{dit}(\pi\vee
\sigma)=\operatorname*{dit}\left(  \pi\right)  \cup\operatorname*{dit}\left(
\sigma\right)  $, so that union is never a disjoint union (when the dit sets
are non-empty). We have used the motivation of thinking of a
partition-as-dit-set $\operatorname*{dit}\left(  \pi\right)  $ as an
\textquotedblleft event\textquotedblright\ in a sample space $U\times U$ with
the probability of that event being the logical entropy of the partition. The
following proposition shows that this motivation extends to the notion of independence.

\begin{proposition}
If $\pi$ and $\sigma$ are (stochastically) independent partitions, then their
dit sets $\operatorname*{dit}\left(  \pi\right)  $ and $\operatorname*{dit}%
\left(  \sigma\right)  $ are independent as events in the sample space
$U\times U$ (with equiprobable points).
\end{proposition}

\noindent For independent partitions $\pi$ and $\sigma$, we need to show that
the probability $m(\pi,\sigma)$ of the event $\operatorname*{Mut}\left(
\pi,\sigma\right)  =\operatorname*{dit}\left(  \pi\right)  \cap
\operatorname*{dit}\left(  \sigma\right)  $ is equal to the product of the
probabilities $h\left(  \pi\right)  $ and $h\left(  \sigma\right)  $ of the
events $\operatorname*{dit}\left(  \pi\right)  $ and $\operatorname*{dit}%
\left(  \sigma\right)  $ in the sample space $U\times U$. By the assumption of
independence, we have $\frac{\left\vert B\cap C\right\vert }{\left\vert
U\right\vert }=p_{B\cap C}=p_{B}p_{C}=\frac{\left\vert B\right\vert \left\vert
C\right\vert }{\left\vert U\right\vert ^{2}}$ so that $\left\vert B\cap
C\right\vert =\left\vert B\right\vert \left\vert C\right\vert /\left\vert
U\right\vert $. By the previous structure theorem for the mutual information
set: $\operatorname*{Mut}\left(  \pi,\sigma\right)  =%
{\textstyle\bigcup\limits_{B\in\pi,C\in\sigma}}
\left(  B-\left(  B\cap C\right)  \right)  \times\left(  C-\left(  B\cap
C\right)  \right)  $, where the union is disjoint so that:

\begin{center}%
\begin{align*}
\left\vert \operatorname*{Mut}\left(  \pi,\sigma\right)  \right\vert  &  =%
{\displaystyle\sum\limits_{B\in\pi,C\in\sigma}}
\left(  \left\vert B\right\vert -\left\vert B\cap C\right\vert \right)
\left(  \left\vert C\right\vert -\left\vert B\cap C\right\vert \right) \\
&  =%
{\displaystyle\sum\limits_{B\in\pi,C\in\sigma}}
\left(  \left\vert B\right\vert -\frac{|B|\left\vert C\right\vert }{\left\vert
U\right\vert }\right)  \left(  \left\vert C\right\vert -\frac{|B|\left\vert
C\right\vert }{\left\vert U\right\vert }\right) \\
&  =\frac{1}{\left\vert U\right\vert ^{2}}%
{\displaystyle\sum\limits_{B\in\pi,C\in\sigma}}
\left\vert B\right\vert \left(  \left\vert U\right\vert -\left\vert
C\right\vert \right)  \left\vert C\right\vert \left(  \left\vert U\right\vert
-\left\vert B\right\vert \right) \\
&  =\frac{1}{\left\vert U\right\vert ^{2}}%
{\displaystyle\sum\limits_{B\in\pi}}
\left\vert B\right\vert \left\vert U-B\right\vert
{\displaystyle\sum\limits_{C\in\sigma}}
\left\vert C\right\vert \left\vert U-C\right\vert \\
&  =\frac{1}{\left\vert U\right\vert ^{2}}\left\vert \operatorname*{dit}%
\left(  \pi\right)  \right\vert \left\vert \operatorname*{dit}\left(
\sigma\right)  \right\vert .
\end{align*}

\end{center}

\noindent Hence under independence, the normalized dit count $m(\pi
,\sigma)=\frac{\left\vert \operatorname*{Mut}\left(  \pi,\sigma\right)
\right\vert }{\left\vert U\right\vert ^{2}}=\frac{\operatorname*{dit}\left(
\pi\right)  }{\left\vert U\right\vert ^{2}}\frac{\operatorname*{dit}\left(
\sigma\right)  }{\left\vert U\right\vert ^{2}}=h\left(  \pi\right)  h\left(
\sigma\right)  $ of the mutual information set $\operatorname*{Mut}\left(
\pi,\sigma\right)  =\operatorname*{dit}\left(  \pi\right)  \cap
\operatorname*{dit}\left(  \sigma\right)  $ is equal to product of the
normalized dit counts of the partitions:

\begin{center}
$m(\pi,\sigma)=h\left(  \pi\right)  h\left(  \sigma\right)  $ if $\pi$ and
$\sigma$ are independent.$\blacksquare$
\end{center}

\subsection{Mutual Information}

For each of the major concepts in the information theory based on the usual
Shannon measure, there should be a corresponding concept based on the
normalized dit counts of logical entropy.\footnote{See Cover and Thomas' book
\cite{cover:eit} for more background on the standard concepts. The
corresponding notions for the block-count entropy are obtained from the usual
Shannon entropy notions by taking antilogs.} In the following sections, we
give some of these corresponding concepts and results.

The logical mutual information of two partitions $m\left(  \pi,\sigma\right)
$ is the normalized dit count of the intersection of their dit-sets:

\begin{center}
$m\left(  \pi,\sigma\right)  =\frac{\left\vert \operatorname*{dit}\left(
\pi\right)  \cap\operatorname*{dit}\left(  \sigma\right)  \right\vert
}{\left\vert U\times U\right\vert }$.
\end{center}

\noindent For Shannon's notion\ of mutual information, we might apply the Venn
diagram heuristics using a block $B\in\pi$ and a block $C\in\sigma$. We saw
before that the information contained in a block $B$ was $H\left(  B\right)
=\log\left(  \frac{1}{p_{B}}\right)  $ and similarly for $C$ while $H\left(
B\cap C\right)  =\log\left(  \frac{1}{p_{B\cap C}}\right)  $ would correspond
to the union of the information in $B$ and in $C$. Hence the overlap or
\textquotedblleft mutual information\textquotedblright\ in $B$ and $C$ could
be motivated as the sum of the two informations minus the union:

\begin{center}
$I\left(  B;C\right)  =\log\left(  \frac{1}{p_{B}}\right)  +\log\left(
\frac{1}{p_{C}}\right)  -\log\left(  \frac{1}{p_{B\cap C}}\right)
=\log\left(  \frac{1}{p_{B}p_{C}}\right)  +\log\left(  p_{B\cap C}\right)
=\log\left(  \frac{p_{B\cap C}}{p_{B}p_{C}}\right)  $.
\end{center}

\noindent Then the (Shannon) \textit{mutual information} in the two partitions
is obtained by averaging over the mutual information for each pair of blocks
from the two partitions:

\begin{center}
$I\left(  \pi;\sigma\right)  =\sum_{B,C}p_{B\cap C}\log\left(  \frac{p_{B\cap
C}}{p_{B}p_{C}}\right)  $.
\end{center}

\noindent The mutual information can be expanded to verify the Venn diagram heuristics:

\begin{center}
$I\left(  \pi;\sigma\right)  =\sum_{B\in\pi,C\in\sigma}p_{B\cap C}\log\left(
\frac{p_{B\cap C}}{p_{B}p_{C}}\right)  =\sum_{B,C}p_{B\cap C}\log\left(
p_{B\cap C}\right)  +\sum_{B,C}p_{B\cap C}\log\left(  \frac{1}{p_{B}}\right)
+\sum_{B,C}p_{B\cap C}\log\left(  \frac{1}{p_{C}}\right)  $

$=-H\left(  \pi\vee\sigma\right)  +\sum_{B\in\pi}p_{B}\log\left(  \frac
{1}{p_{B}}\right)  +\sum_{C\in\sigma}p_{C}\log\left(  \frac{1}{p_{C}}\right)
=H\left(  \pi\right)  +H\left(  \sigma\right)  -H\left(  \pi\vee\sigma\right)
$.
\end{center}

\noindent We will later see an important inequality, $I\left(  \pi
;\sigma\right)  \geq0$ (with equality under independence), and its logical version.

In the logical theory, the corresponding \textquotedblleft modular
law\textquotedblright\ follows from the inclusion-exclusion principle applied
to dit-sets: $\left\vert \operatorname*{dit}\left(  \pi\right)  \cap
\operatorname*{dit}\left(  \sigma\right)  \right\vert =\left\vert
\operatorname*{dit}\left(  \pi\right)  \right\vert +\left\vert
\operatorname*{dit}\left(  \sigma\right)  \right\vert -\left\vert
\operatorname*{dit}\left(  \pi\right)  \cup\operatorname*{dit}\left(
\sigma\right)  \right\vert $. Normalizing yields:

\begin{center}
$m\left(  \pi,\sigma\right)  =\frac{\left\vert \operatorname*{dit}\left(
\pi\right)  \cap\operatorname*{dit}\left(  \sigma\right)  \right\vert
}{\left\vert U\right\vert ^{2}}=\frac{\left\vert \operatorname*{dit}\left(
\pi\right)  \right\vert }{\left\vert U\right\vert ^{2}}+\frac{\left\vert
\operatorname*{dit}\left(  \sigma\right)  \right\vert }{\left\vert
U\right\vert ^{2}}-\frac{\left\vert \operatorname*{dit}\left(  \pi\right)
\cup\operatorname*{dit}\left(  \sigma\right)  \right\vert }{\left\vert
U\right\vert ^{2}}=h\left(  \pi\right)  +h\left(  \sigma\right)  -h\left(
\pi\vee\sigma\right)  $.
\end{center}

Since the formulas concerning the logical and Shannon entropies often have
similar relationships, e.g., $I\left(  \pi;\sigma\right)  =H\left(
\pi\right)  +H\left(  \sigma\right)  -H\left(  \pi\vee\sigma\right)  $ and
$m\left(  \pi,\sigma\right)  =h\left(  \pi\right)  +h\left(  \sigma\right)
-h\left(  \pi\vee\sigma\right)  $, it is useful to also emphasize some crucial
differences. One of the most important special cases is for two partitions
that are (stochastically) independent. For independent partitions, it is
immediate that $I\left(  \pi;\sigma\right)  =\sum_{B,C}p_{B\cap C}\log\left(
\frac{p_{B\cap C}}{p_{B}p_{C}}\right)  =0$ but we have already seen that for
the logical mutual information, $m\left(  \pi,\sigma\right)  >0$ so long as
neither partition is the blob $\widehat{0} $. However for independent
partitions we have;

\begin{center}
$m\left(  \pi,\sigma\right)  =h\left(  \pi\right)  h\left(  \sigma\right)  $
\end{center}

\noindent so the logical mutual information behaves like the probability of
both events occurring in the case of independence (as it must since logical
entropy concepts have direct probabilistic interpretations). For independent
partitions, the relation $m\left(  \pi,\sigma\right)  =h\left(  \pi\right)
h\left(  \sigma\right)  $ means that the probability that a random pair is
distinguished by both partitions is the same as the probability that it is
distinguished by one partition \textit{times} the probability that it is
distinguished by the other partition. In simpler terms, for independent $\pi$
and $\sigma$, the probability that $\pi$ and $\sigma$ distinguishes is the
probability that $\pi$ distinguishes times the probability that $\sigma$ distinguishes.

It is sometimes convenient to think in the complementary terms of an
equivalence relation \textquotedblleft identifying.\textquotedblright\ rather
than a partition distinguishing. Since $h\left(  \pi\right)  $ can be
interpreted as the probability that a random pair of elements from $U$ are
distinguished by $\pi$, i.e., as a distinction probability, its complement
$1-h\left(  \pi\right)  $ can be interpreted as an \textit{identification
probability}, i.e., the probability that a random pair is identified by $\pi$
(thinking of $\pi$ as an equivalence relation on $U$). In general,

\begin{center}
$\left[  1-h\left(  \pi\right)  \right]  \left[  1-h\left(  \sigma\right)
\right]  =1-h\left(  \pi\right)  -h\left(  \sigma\right)  +h\left(
\pi\right)  h\left(  \sigma\right)  =\left[  1-h\left(  \pi\vee\sigma\right)
\right]  +\left[  h\left(  \pi\right)  h\left(  \sigma\right)  -m(\pi
,\sigma\right]  $
\end{center}

\noindent which could also be rewritten as:

\begin{center}
$\left[  1-h\left(  \pi\vee\sigma\right)  \right]  -\left[  1-h\left(
\pi\right)  \right]  \left[  1-h\left(  \sigma\right)  \right]  =m(\pi
,\sigma)-h\left(  \pi\right)  h\left(  \sigma\right)  $.
\end{center}

\noindent Hence:

\begin{center}
if $\pi$ and $\sigma$ are independent: $\left[  1-h\left(  \pi\right)
\right]  \left[  1-h\left(  \sigma\right)  \right]  =\left[  1-h\left(
\pi\vee\sigma\right)  \right]  $.
\end{center}

\noindent Thus if $\pi$ and $\sigma$ are independent, then the probability
that the join partition $\pi\vee\sigma$ identifies is the probability that
$\pi$ identifies times the probability that $\sigma$ identifies. In summary,
if $\pi$ and $\sigma$ are independent, then:%

\begin{align*}
\text{Binary-partition-count (Shannon) entropy}  &  \text{:}\text{ }H\left(
\pi\vee\sigma\right)  =H\left(  \pi\right)  +H\left(  \sigma\right) \\
\text{Block-count entropy}  &  \text{:}\text{ }H_{m}\left(  \pi\vee
\sigma\right)  =H_{m}\left(  \pi\right)  H_{m}\left(  \sigma\right) \\
\text{Normalized-dit-count (logical) entropy}  &  \text{:}\text{ }h\left(
\pi\vee\sigma\right)  =1-\left[  1-h\left(  \pi\right)  \right]  \left[
1-h\left(  \sigma\right)  \right]  .
\end{align*}

\subsection{Cross Entropy and Divergence}

Given a set partition $\pi=\left\{  B\right\}  _{B\in\pi}$ on a set $U$, the
\textquotedblleft natural\textquotedblright\ or Laplacian probability
distribution on the blocks of the partition was $p_{B}=\frac{\left\vert
B\right\vert }{\left\vert U\right\vert }$. The set partition $\pi$ also
determines the set of distinctions $\operatorname*{dit}\left(  \pi\right)
\subseteq U\times U$ and the logical entropy of the partition was the
Laplacian probability of the dit-set as an event, i.e., $h\left(  \pi\right)
=\frac{\left\vert \operatorname*{dit}(\pi\right\vert }{\left\vert U\times
U\right\vert }=\sum_{B}p_{B}\left(  1-p_{B}\right)  $. But we may also
\textquotedblleft kick away the ladder\textquotedblright\ and generalize all
the definitions to any finite probability distributions $p=\left\{
p_{1},...,p_{n}\right\}  $. A probability distribution $p$ might be given by
finite-valued random variables $X$ on a sample space $U$ where $p_{i}%
=\operatorname*{Prob}(X=x_{i})$ for the finite set of distinct values $x_{i}$
for $i=1,...,n$. Thus the logical entropy of the random variable $X$ is:
$h\left(  X\right)  =\sum_{i=1}^{n}p_{i}\left(  1-p_{i}\right)  =1-\sum
_{i}p_{i}^{2}$. The entropy is only a function of the probability distribution
of the random variable, not its values, so we could also take it simply as a
function of the probability distribution $p$, $h\left(  p\right)  =1-\sum
_{i}p_{i}^{2}$. Taking the sample space as $\left\{  1,...,n\right\}  $, the
logical entropy is still interpreted as the probability that two independent
draws will draw distinct points from $\left\{  1,...,n\right\}  $. The further
generalizations replacing probabilities by probability density functions and
sums by integrals are straightforward but beyond the scope of this paper
(which is focused on conceptual foundations rather than mathematical developments).

Given two probability distributions $p=\left\{  p_{1},...,p_{n}\right\}  $ and
$q=\left\{  q_{1},...,q_{n}\right\}  $ on the same sample space $\left\{
1,...,n\right\}  $, we can again consider the drawing of a pair of points but
where the first drawing is according to $p$ and the second drawing according
to $q$. The probability that the pair of points is distinct would be a natural
and more general notion of logical entropy which we will call the:

\begin{center}
\textit{logical} \textit{cross entropy: }$h\left(  p\Vert q\right)  =\sum
_{i}p_{i}(1-q_{i})=1-\sum_{i}p_{i}q_{i}=\sum_{i}q_{i}(1-p_{i})=h\left(  q\Vert
p\right)  $
\end{center}

\noindent which is symmetric. The logical cross entropy is the same as the
logical entropy when the distributions are the same, i.e.,

\begin{center}
if $p=q$, then $h\left(  p\Vert q\right)  =h\left(  p\right)  $.
\end{center}

The notion of \textit{cross entropy} in conventional information theory is:
$H\left(  p\Vert q\right)  =\sum_{i}p_{i}\log\left(  \frac{1}{q_{i}}\right)  $
which is not symmetrical due to the asymmetric role of the logarithm, although
if $p=q$, then $H\left(  p\Vert q\right)  =H\left(  p\right)  $. Then the
\textit{Kullback-Leibler divergence} $D\left(  p\Vert q\right)  =\sum_{i}%
p_{i}\log\left(  \frac{p_{i}}{q_{i}}\right)  $ is defined as a measure of the
distance or divergence between the two distributions where $D\left(  p\Vert
q\right)  =H\left(  p\Vert q\right)  -H\left(  p\right)  $. The
\textit{information inequality}\ is: $D\left(  p\Vert q\right)  \geq0$ with
equality if and only if $p_{i}=q_{i}$ for $i=1,...,n$ \cite[p. 26]{cover:eit}.
Given two partitions $\pi$ and $\sigma$, the inequality $I\left(  \pi
;\sigma\right)  \geq0$ is obtained by applying the information inequality to
the two distributions $\left\{  p_{B\cap C}\right\}  $ and $\left\{
p_{B}p_{C}\right\}  $ on the sample space $\left\{  \left(  B,C\right)
:B\in\pi,C\in\sigma\right\}  =\pi\times\sigma$:

\begin{center}
$I\left(  \pi;\sigma\right)  =\sum_{B,C}p_{B\cap C}\log\left(  \frac{p_{B\cap
C}}{p_{B}p_{C}}\right)  =D\left(  \left\{  p_{B\cap C}\right\}  \Vert\left\{
p_{B}p_{C}\right\}  \right)  \geq0$ with equality under independence.
\end{center}

But starting afresh, one might ask: \textquotedblleft What is the natural
measure of the difference or distance between two probability distributions
$p=\left\{  p_{1},...,p_{n}\right\}  $ and $q=\left\{  q_{1},...,q_{n}%
\right\}  $ that would always be non-negative, and would be zero if and only
they are equal?\textquotedblright\ The (Euclidean) distance between the two
points in $%
\mathbb{R}
^{n}$ would seem to be the \textquotedblleft logical\textquotedblright%
\ answer---so we take that distance (squared) as the definition of the:

\begin{center}
\textit{logical divergence} (or \textit{logical} \textit{relative entropy}):
$d\left(  p\Vert q\right)  =$ $\sum_{i}\left(  p_{i}-q_{i}\right)  ^{2}$,
\end{center}

\noindent which is symmetric and non-negative. We have component-wise:

\begin{center}
$0\leq\left(  p_{i}-q_{i}\right)  ^{2}=p_{i}^{2}-2p_{i}q_{i}+q_{i}%
^{2}=2\left[  \frac{1}{n}-p_{i}q_{i}\right]  -\left[  \frac{1}{n}-p_{i}%
^{2}\right]  -\left[  \frac{1}{n}-q_{i}^{2}\right]  $
\end{center}

\noindent so that taking the sum for $i=1,...,n$ gives:

\begin{center}
$0\leq d\left(  p\Vert q\right)  =\sum_{i}\left(  p_{i}-q_{i}\right)
^{2}=2\left[  1-\sum_{i}p_{i}q_{i}\right]  -\left[  1-\sum_{i}p_{i}%
^{2}\right]  -\left[  1-\sum_{i}q_{i}^{2}\right]  =2h\left(  p\Vert q\right)
-h\left(  p\right)  -h\left(  q\right)  $.
\end{center}

\noindent Thus we have the:

\begin{center}
\fbox{\textit{\ }$0\leq d\left(  p\Vert q\right)  =2h\left(  p\Vert q\right)
-h\left(  p\right)  -h\left(  q\right)  $ with equality if and only if
$p_{i}=q_{i}$ for $i=1,...,n$}

Logical information inequality.
\end{center}

\noindent If we take $h\left(  p\Vert q\right)  -\frac{1}{2}\left[  h\left(
p\right)  +h\left(  q\right)  \right]  $ as the \textit{Jensen difference}%
\ \cite[p. 25]{rao:div} between the two distributions, then the logical
divergence is twice the Jensen difference. The half-and-half probability
distribution $\frac{p+q}{2}$ that mixes $p$ and $q$ has the logical entropy of
$h\left(  \frac{p+q}{2}\right)  =\frac{h\left(  p\Vert q\right)  }{2}%
+\frac{h\left(  p\right)  +h\left(  q\right)  }{4}$ so that:

\begin{center}
$d\left(  p\Vert q\right)  =4\left[  h\left(  \frac{p+q}{2}\right)  -\frac
{1}{2}\left\{  h\left(  p\right)  +h\left(  q\right)  \right\}  \right]
\geq0$.
\end{center}

\noindent The logical information inequality tells us that \textquotedblleft
mixing increases logical entropy\textquotedblright\ (or, to be precise, mixing
does not decrease logical entropy) which also follows from the fact that
logical entropy $h\left(  p\right)  =1-\sum_{i}p_{i}^{2}$ is a concave function.

An important special case of the logical information inequality is when
$p=\left\{  p_{1},...,p_{n}\right\}  $ is the uniform distribution with all
$p_{i}=\frac{1}{n}$. Then $h\left(  p\right)  =1-\frac{1}{n}$ where the
probability that a random pair is distinguished (i.e., the random variable $X
$ with $\operatorname*{Prob}(X=x_{i})=p_{i}$ has different values in two
independent samples) takes the specific form of the probability $1-\frac{1}%
{n}$ that the second draw gets a different value than the first. It may at
first seem counterintuitive that in this case the cross entropy is $h(p\Vert
q)=h\left(  p\right)  +\sum_{i}p_{i}\left(  p_{i}-q_{i}\right)  =h\left(
p\right)  +\sum_{i}\frac{1}{n}\left(  \frac{1}{n}-q_{i}\right)  =h\left(
p\right)  =1-\frac{1}{n}$ for \textit{any} $q=\left\{  q_{1},...,q_{n}%
\right\}  $. But $h\left(  p\Vert q\right)  $ is the probability that the two
points, say $i$ and $i^{\prime}$, in the sample space $\left\{
1,...,n\right\}  $ are distinct when one draw was according to $p$ and the
other according to $q $. Taking the first draw according to $q$, the
probability that the second draw is distinct from whatever point was
determined in the first draw is indeed $1-\frac{1}{n}$ (regardless of
probability $q_{i}$ of the point drawn on the first draw). Then the divergence
$d\left(  p\Vert q\right)  =2h\left(  p\Vert q\right)  -h\left(  p\right)
-h\left(  q\right)  =\left(  1-\frac{1}{n}\right)  -h\left(  q\right)  $ is a
non-negative measure of how much the probability distribution $q$ diverges
from the uniform distribution. It is simply the difference in the probability
that a random pair will be distinguished by the uniform distribution and by
$q$. Also since $0\leq d\left(  p\Vert q\right)  $, this shows that among all
probability distributions on $\left\{  1,...,n\right\}  $, the uniform
distribution has the maximum logical entropy. In terms of partitions, the
$n$-block partition with $p_{B}=\frac{1}{n}$ has maximum logical entropy among
all $n$-block partitions. In the case of $\left\vert U\right\vert $ divisible
by $n$, the equal $n$-block partitions make more distinctions than any of the
unequal $n$-block partitions on $U$.

For any partition $\pi$ with the $n$ block probabilities $\left\{
p_{B}\right\}  _{B\in\pi}=\left\{  p_{1},...,p_{n}\right\}  $:

\begin{center}
$h\left(  \pi\right)  \leq1-\frac{1}{n}$ with equality if and only if
$p_{1}=...=p_{n}=\frac{1}{n}$.
\end{center}

\noindent For the corresponding results in the Shannon's information theory,
we can apply the information inequality $D\left(  p\Vert q\right)  =H\left(
p\Vert q\right)  -H\left(  p\right)  \geq0$ with $q$ as the uniform
distribution $q_{1}=...=q_{n}=\frac{1}{n}$. Then $H\left(  p\Vert q\right)
=\sum_{i}p_{i}\log\left(  \frac{1}{1/n}\right)  =\log\left(  n\right)  $ so
that: $H(p)\leq\log\left(  n\right)  $ or in terms of partitions:

\begin{center}
$H\left(  \pi\right)  \leq\log_{2}\left(  \left\vert \pi\right\vert \right)  $
with equality if and only if the probabilities are equal
\end{center}

\noindent or, in base-free terms,

\begin{center}
$H_{m}\left(  \pi\right)  \leq\left\vert \pi\right\vert $ with equality if and
only if the probabilities are equal.
\end{center}

\noindent The three entropies take their maximum values (for fixed number of
blocks $\left\vert \pi\right\vert $) at the partitions with equiprobable blocks.

In information theory texts, it is customary to graph the case of $n=2$ where
the entropy is graphed as a function of $p_{1}=p$ with $p_{2}=1-p$. The
Shannon entropy function $H\left(  p\right)  =-p\log\left(  p\right)  -\left(
1-p\right)  \log\left(  1-p\right)  $ looks somewhat like an inverted parabola
with its maximum value of $\log(n)=\log\left(  2\right)  =1$ at $p=.5$. The
logical entropy function $h\left(  p\right)  =1-p^{2}-\left(  1-p\right)
^{2}=2p-2p^{2}=2p\left(  1-p\right)  $ is an inverted parabola with its
maximum value of $1-\frac{1}{n}=1-\frac{1}{2}=.5$ at $p=.5$. \noindent The
block-count entropy $H_{m}\left(  p\right)  =\left(  \frac{1}{p}\right)
^{p}\left(  \frac{1}{1-p}\right)  ^{1-p}=2^{H\left(  p\right)  }$ is an
inverted U-shaped curve that starts and ends at $1=2^{H(0)}=2^{H\left(
1\right)  }$ and has its maximum at $2=2^{H\left(  .5\right)  }$.

\subsection{Summary of Analogous Concepts and Results}

\begin{center}%
\begin{tabular}
[c]{l|c|c|}\cline{2-3}
& $\text{Shannon Entropy}$ & $\text{Logical Entropy}$\\\hline
\multicolumn{1}{|l|}{$\text{Block Entropy}$} & $H\left(  B\right)
=\log\left(  1/p_{B}\right)  $ & $h\left(  B\right)  =1-p_{B}$\\\hline
\multicolumn{1}{|l|}{$\text{Relationship}$} & $H\left(  B\right)  =\log\left(
\frac{1}{1-h\left(  B\right)  }\right)  $ & $h\left(  B\right)  =1-\frac
{1}{2^{H\left(  B\right)  }}$\\\hline
\multicolumn{1}{|l|}{$\text{Entropy}$} & $H(\pi)=\sum p_{B}\log\left(
1/p_{B}\right)  $ & $h\left(  \pi\right)  =\sum p_{B}\left(  1-p_{B}\right)
$\\\hline
\multicolumn{1}{|l|}{$\text{Mutual Information}$} & $I(\pi;\sigma)=H\left(
\pi\right)  +H\left(  \sigma\right)  -H\left(  \pi\vee\sigma\right)  $ &
$m\left(  \pi,\sigma\right)  =h\left(  \pi\right)  +h\left(  \sigma\right)
-h\left(  \pi\vee\sigma\right)  $\\\hline
\multicolumn{1}{|l|}{Independence} & $I\left(  \pi;\sigma\right)  =0$ &
$m\left(  \pi,\sigma\right)  =h\left(  \pi\right)  h\left(  \sigma\right)
$\\\hline
\multicolumn{1}{|l|}{Independence \& Joins} & $H\left(  \pi\vee\sigma\right)
=H\left(  \pi\right)  +H\left(  \sigma\right)  $ & $h\left(  \pi\vee
\sigma\right)  =1-\left[  1-h\left(  \pi\right)  \right]  \left[  1-h\left(
\sigma\right)  \right]  $\\\hline
\multicolumn{1}{|l|}{$\text{Cross Entropy}$} & $H\left(  p\Vert q\right)
=\sum p_{i}\log\left(  1/q_{i}\right)  $ & $h\left(  p\Vert q\right)  =\sum
p_{i}\left(  1-q_{i}\right)  $\\\hline
\multicolumn{1}{|l|}{$\text{Divergence}$} & $D\left(  p\Vert q\right)
=H\left(  p\Vert q\right)  -H\left(  p\right)  $ & $d\left(  p\Vert q\right)
=2h\left(  p\Vert q\right)  -h\left(  p\right)  -h\left(  q\right)  $\\\hline
\multicolumn{1}{|l|}{$\text{Information Inequality}$} & $D\left(  p\Vert
q\right)  \geq0\text{ with }=\text{ iff }p_{i}=q_{i}\forall i$ & $d\left(
p\Vert q\right)  \geq0\text{ with }=\text{ iff }p_{i}=q_{i}\forall i$\\\hline
\multicolumn{1}{|l|}{$\text{Info.\ Ineq. }$Sp. Case} & $I\left(  \pi
;\sigma\right)  =D\left(  \left\{  p_{B\cap C}\right\}  \Vert\left\{
p_{B}p_{C}\right\}  \right)  \geq0$ & $d\left(  \left\{  p_{B\cap C}\right\}
\Vert\left\{  p_{B}p_{C}\right\}  \right)  \geq0$\\\cline{1-1}
& $\text{with equality under independence}$ & $\text{with equality under
independence.}$\\\cline{2-3}%
\end{tabular}

\end{center}

\section{Concluding Remarks}

In the duality of subsets of a set with partitions on a set, we found that the
elements of a subset were dual to the distinctions (dits) of a partition. Just
as the finite probability theory for events started by taking the size of a
subset (\textquotedblleft event\textquotedblright) $S$ normalized to the size
of the finite universe $U$ as the probability $\operatorname*{Prob}\left(
S\right)  =\frac{\left\vert S\right\vert }{\left\vert U\right\vert }$, so it
would be natural to consider the corresponding theory that would associate
with a partition $\pi$ on a finite $U$, the size $\left\vert
\operatorname*{dit}\left(  \pi\right)  \right\vert $ of the set of
distinctions of the partition normalized by the total number of ordered pairs
$\left\vert U\times U\right\vert $. This number $h\left(  \pi\right)
=\frac{\left\vert \operatorname*{dit}\left(  \pi\right)  \right\vert
}{\left\vert U\times U\right\vert }$ was called the logical entropy of $\pi$
and could be interpreted as the probability that a randomly picked (with
replacement) pair of elements from $U$ is distinguished by the partition $\pi
$, just as $\operatorname*{Prob}\left(  s\right)  =\frac{\left\vert
S\right\vert }{\left\vert U\right\vert }$ is the probability that a randomly
picked element from $U$ is an element of the subset $S$. Hence this notion of
logical entropy arises naturally out of the logic of partitions that is dual
to the usual logic of subsets.

The question immediately arises of the relationship with Shannon's concept of
entropy. Following Shannon's definition of entropy, there has been a veritable
plethora of suggested alternative entropy concepts \cite{kapur:mi}. Logical
entropy is \textit{not} an alternative entropy concept intended to displace
Shannon's concept any more than is the block-count entropy concept. Instead, I
have argued that the dit-count, block-count, and binary-partition-count
concepts of entropy should be seen as three ways to measure that same
\textquotedblleft information\textquotedblright\ expressed in its most atomic
terms as distinctions.\ The block-count entropy, although it can be
independently defined, is trivially related to Shannon's
binary-partition-count concept---just take antilogs. The relationship of the
logical concept of entropy to the Shannon concept is a little more subtle but
is quite simple at the level of blocks $B\in\pi$: $h\left(  B\right)
=1-p_{B}$, $H_{m}\left(  B\right)  =\frac{1}{p_{B}}$, and $H\left(  B\right)
=\log\left(  \frac{1}{p_{B}}\right)  $ so that eliminating the probability, we have:%

\begin{align*}
h\left(  B\right)   &  =1-\frac{1}{H_{m}\left(  B\right)  }\\
&  =1-\frac{1}{2^{H\left(  B\right)  }}.
\end{align*}

\noindent Then the logical and additive entropies for the whole partition are
obtained by taking the (additive) expectation of the block entropies while the
block-count entropy is the multiplicative expectation of the block entropies:%

\begin{align*}
H_{m}\left(  \pi\right)   &  =\prod_{B\in\pi}\left(  \frac{1}{p_{B}}\right)
^{p_{B}}\\
H\left(  \pi\right)   &  =\sum_{B\in\pi}p_{B}\log\left(  \frac{1}{p_{B}%
}\right) \\
h\left(  \pi\right)   &  =\sum_{B\in\pi}p_{B}\left(  1-p_{B}\right)  .
\end{align*}

In conclusion, the simple root of the matter is three different ways to
\textquotedblleft measure\textquotedblright\ the distinctions that generate an
$n$-element set. Consider a $4$ element set. One measure of the distinctions
that distinguish a set of $4$ elements is its cardinality $4$, and that
measure leads to the block-count entropy. Another measure of that set is
$\log_{2}(4)=2$ which can be interpreted as the minimal number of binary
partitions necessary: (a) to single out any designated element as a singleton
(search interpretation) or, equivalently, (b) to distinguish all the elements
from each other (distinction interpretation). That measure leads to Shannon's
entropy formula. And the third measure is the (normalized) count of
distinctions (counted as ordered pairs) necessary to distinguish all the
elements from each other, i.e., $\frac{4\times4-4}{4\times4}=\frac{12}%
{16}=\frac{3}{4}$, which yields the logical entropy formula. These measures
stand in the block value relationship: $\frac{3}{4}=1-\frac{1}{4}=1-\frac
{1}{2^{2}}$. It is just a matter of:

\begin{enumerate}
\item counting the elements distinguished (block-count entropy),

\item counting the binary partitions needed to distinguish the elements
(Shannon entropy), or

\item counting the (normalized) distinctions themselves (logical entropy).
\end{enumerate}

\end{document}